\documentclass{ar-1col-S2O}
\usepackage[numbers]{natbib}
\usepackage{url}
\usepackage{amssymb}
\usepackage[colorlinks = red, linkcolor=red, anchorcolor=black, citecolor=blue]{hyperref}

\jname{Ann. Rev. Nucl. Part. Sci.}
\jvol{75}
\jyear{2025}
\doi{10.1146/annurev-nucl-121423-100931}

\newcommand{\CC}[2]{\mathbb{C}^{#1}_{#2}}
\newcommand{\V}{V^\prime}

\newcommand{\rS}{{\rm S}}

\newcommand{\rV}{{\rm V}}

\newcommand{\rX}{{\rm X}}

\newcommand{\rD}{{\rm D}}

\newcommand{\GeV}{\rm \, GeV}

\newcommand{\keV}{\rm \, keV}

\definecolor{coral}{RGB}{255,127,80}

\begin{document}

\markboth{Martin Camalich - Ziegler}{ Flavor phenomenology of light dark sectors}

\title{Flavor phenomenology of light dark sectors}

\author{Jorge Martin Camalich,$^{1,2}$ and Robert Ziegler$^{3}$
\affil{$^1$Instituto de Astrof\'isica de Canarias,  C/ V\'ia L\'actea, s/n E38205 - La Laguna, Tenerife, Spain}
\affil{$^2$Universidad de La Laguna, Departamento de Astrof\'isica, La Laguna, Tenerife, Spain}
\affil{$^3$Institute for Theoretical Particle Physics, Karlsruhe Institute of Technology, 76128 Karlsruhe, Germany}}

\begin{abstract}
The dark sector offers a compelling theoretical framework for addressing the nature of dark matter while potentially solving other fundamental problems in physics. This review focuses on light dark  \textit{flavored} sector models, which are those where the flavor structure of the interactions with the standard model is non-trivial and distinguish among different fermion families. Such scenarios feature flavor violation leading to unique experimental signatures, such as flavor-changing neutral current decays of heavy hadrons (kaons, $D$ and $B$ mesons, baryons) and leptons (muons and taus) with missing energy carried away by light dark-sector particles. In this article, we review their motivation, summarize current constraints, highlight discovery opportunities in ongoing and future flavor experiments, and discuss implications for astrophysics and cosmology.
\end{abstract}

\maketitle

\tableofcontents

\section{INTRODUCTION}

Despite the remarkable success of the standard model of particle physics (SM) and the standard cosmological model ($\Lambda$CDM), several fundamental questions remain unanswered. What is the nature of dark matter and dark energy? What caused the matter-antimatter asymmetry observed in the universe?  Why is charge-parity conserved in the strong interactions? What is the source of flavor and why does matter exhibit three distinct families? These and other questions continue to drive extensive theoretical and experimental efforts aimed at uncovering new laws of nature and potentially revealing entirely new sectors of constituents that can address these foundational problems.

One general theoretical framework that primarily addresses the nature of dark matter (DM), while potentially solving other fundamental questions, is the ``dark sector''. This can describe a minimal scenario involving a new particle, such as the QCD axion, weakly coupled with the SM and playing the role of DM. More broadly, it often refers to an extended sector comprising several new particles and interactions. Over the past decade, there has been a surge of theoretical activity in characterizing these dark sectors, especially exploring their connections to other fundamental puzzles, and driving novel ideas for their detection and potential discovery through experiments and observations~\cite{Knapen:2017xzo,Lin:2019uvt,Zurek:2024qfm,Cirelli:2024ssz}.

In this review, we focus on a specific aspect of the dark sector: the flavor structure of their interactions with the SM. In particular, we examine cases where this flavor structure is \textit{non-trivial} (i.e., not proportional to the identity matrix) in generation space, but where the interactions and couplings with the dark sector distinguish among the fermion families of the SM. Such scenarios naturally arise in dark sectors intertwined with solutions to the SM flavor puzzle or baryogenesis. The prototypical example is the QCD axion with non-diagonal flavor couplings~\cite{Wilczek:1982rv}, which simultaneously addresses the DM, strong CP, and flavor problems. Other examples include the dark photon~\cite{Holdom:1985ag,Dobrescu:2004wz} and dark baryons~\cite{Elor:2018twp}.

One of the most significant implications of these dark-flavored sectors is their potential to induce flavor-violating transitions among SM fermions. Furthermore, if the dark sector particle mediating these transitions is sufficiently light, then it can be produced in heavy-flavor decays and detected through missing energy signatures~\cite{Kamenik:2011vy}. This phenomenology opens up a wide range of unexplored avenues for discovery in flavor experiments, focused on the decays of heavy flavored hadrons, such as  kaons, $D$- or $B$-mesons and baryons, or heavy leptons like muons and taus. These opportunities are especially timely, as flavor physics is currently undergoing a golden age in experimental precision and reach. This includes multipurpose experiments in flavor factories such as LHCb~\cite{LHCb:2018roe}, Belle~II~\cite{Belle-II:2018jsg}, and BESIII~\cite{BESIII:2020nme}, or specialized experiments such as NA62~\cite{NA62:2017rwk}, KOTO~\cite{Yamanaka:2012yma}, or  MEG II \cite{Baldini:2018nnn} that target specific rare decays with exceptional sensitivity. 

In this work, we review the phenomenology of light dark-flavored sectors, where ``light'' refers to particles that are effectively massless or have negligible masses relative to the energy scales of interest - see Sec.~\ref{setmot}. Specifically, we focus on the experimental signatures associated with kinematic configurations of decays involving missing energy and provide an overview of the current constraints on these models derived from flavor-physics experiments - see Sec.~\ref{CL}. Furthermore, we explore the sensitivity and discovery potential of current and future searches specifically designed to target these experimental signatures, highlighting their importance - see Sec.~\ref{FP}. Finally, consequences of dark flavored sectors in astrophysics and cosmology, with the corresponding limits derived on their interactions, are reviewed in Secs.~\ref{AC} and \ref{cosmo}, respectively.


\section{SETUP AND MOTIVATION}
\label{setmot}
 
We explore extensions of the SM by introducing neutral bosonic particles with masses significantly below the GeV scale. Specifically, we consider adding either a new scalar particle $a$ with mass $m_a$ or a new light vector boson $\V_{\mu}$ with mass $m_{\V}$. We start from a basis in which these new states are orthogonal to the SM states; i.e., where a possible kinetic mixing between the photon and the light vector boson has already been diagonalized.

Below the electroweak (EW) scale, the interactions of the new states with the SM fermions can be systematically described by an Effective Field Theory (EFT) approach, by introducing the most general set of operators that respect the unbroken part of the SM gauge group, $SU(3)_c\times U(1)_{\rm em}$. Here, we exclusively focus on flavor-violating interactions, written without loss of generality in the fermion-mass basis. The leading-order interactions, in EFT power counting, of the new bosons are described by the following operators:
\begin{equation}
\label{scalar}
{\cal L}_{\rm scalar}  =  -   \frac{i}{\Lambda} a \, \overline{f}_i \left[ (m_i - m_j) \CC{\rS }{ij} + (m_i + m_j)  \CC{\rS5}{ij}  \gamma_5 \right] f_j   \, ,
\end{equation}
\begin{equation}
\label{vector}
{\cal L}_{\rm vector}  =  \frac{m_{\V}}{\Lambda} \V_\mu \, \overline{f}_i \gamma^\mu \left( \CC{\rV }{ij} + \CC{\rV5}{ij}  \gamma_5 \right)f_j \, ,
\end{equation}
where $i \ne j$ denote SM quark or lepton flavors, and all couplings are hermitian matrices in flavor space, e.g., $(\CC{\rS }{ij})^* =  \CC{\rS }{ji}$.

The couplings are normalized with respect to a UV scale $\Lambda$. For the scalar case we choose a convenient prefactor as discussed below. Note in fact that above the EW scale the couplings in Eq.~\ref{scalar} are not $SU(2)_L$ invariant, and must involve a single power of the EW breaking scale. Furthermore, flavor-violating currents coupled to a vector boson, Eq.~\ref{vector}, are not conserved and must be proportional to at least one power of the $U(1)^\prime$-breaking scale, which is the vector mass $m_{\V}$ upon including the dark gauge coupling, see Ref.~\cite{Eguren:2024oov} for more details. This normalization also ensures finite amplitudes in the $m_{\V} \to 0$ limit, which correspond to the amplitudes with the associated Goldstone bosons as initial or final states. Indeed, in this limit, the longitudinal polarization dominates according to the Goldstone Boson Equivalence theorem. With the replacement $V_\mu^\prime \to \partial_\mu a /m_V^\prime$ and integrating by parts, one recovers the scalar interactions in Eq.~\ref{scalar}, identifying $\CC{\rS }{ij}  = \CC{\rV }{ij}$ and $\CC{\rS5 }{ij}  = \CC{\rV5 }{ij}$. This also justifies the chosen fermion mass normalization factors for scalar couplings in Eq.~\ref{scalar}.

Genuinely new interactions of a very light vector boson can be described by flavor-violating dipole interactions, as given by (see e.g. Ref.~\cite{Fabbrichesi:2017vma})
\begin{equation}
\label{dipole}
{\cal L}_{\rm dipole} =
  \frac{1}{\Lambda} \V_{\mu \nu} \, \overline{f}_i \sigma^{\mu \nu} \left( \CC{\rD }{ij} + i \, \CC{\rD5}{ij} \gamma_5 \right) f_j \, ,
\end{equation}
where $\V_{\mu \nu} = \partial_\mu \V_{\nu} - \partial_\nu \V_{\mu}$, $\sigma^{\mu \nu} = i/2 \, [\gamma^\mu, \gamma^\nu]$,
and the dipole couplings $\CC{\rD }{ij}$ and $\CC{\rD5}{ij}$ are hermitian matrices in flavor space. Note that this operator is naturally of dimension six if the mass scale of the associated UV physics is significantly larger than the electroweak scale, as required by $SU(2)_L$ invariance~\cite{Dobrescu:2004wz}. In this case, the UV scale $\Lambda$ should be replaced by $\Lambda^2/v$, where $v = 174 \GeV$ is the Higgs field vacuum expectation value.

We assume that the other possible interactions with the SM particles, particularly the flavor-diagonal couplings to fermions, are sufficiently
small to ensure that the new bosons remain stable on collider scales. Under this assumption, limits on the flavor-violating couplings of dark bosons described by the  interactions
\begin{equation}
\label{setup}
{\cal L}_{\rm int} = {\cal L}_{\rm scalar} + {\cal L}_{\rm vector} + {\cal L}_{\rm dipole} \, , 
\end{equation}
with Eqs.~\ref{scalar}, \ref{vector} and \ref{dipole} can be derived from hadronic and leptonic decays with missing energy in the final state. These constraints also apply when the dark boson is unstable, but promptly decays into stable invisible particles, for example dark fermions or neutrinos. 

The scenarios described by the above Lagrangians can be well motivated in SM extensions that are capable of (possibly simultaneously) explaining some of the problems and shortcomings of the SM, such as the existence of DM, see e.g~Ref.~\cite{Cirelli:2024ssz}, the absence of CP violation in strong interactions, see e.g.~Ref.~\cite{DiLuzio:2020wdo}, and the flavor puzzle, see e.g.~Ref.~\cite{Feruglio:2015jfa}. In the following, we briefly discuss some of these frameworks. 

\subsection{QCD Axion and light ALP DM}
\label{axion_ALP}

Arguably, the most prominent motivation for the above setup is the QCD axion~\cite{Wilczek:1977pj,Weinberg:1977ma}, which emerges as a low-energy remnant of the Peccei-Quinn (PQ) solution to the strong CP problem~\cite{Peccei:1977hh,Peccei:1977ur}, i.e. the observed absence of CP violation in strong interactions. As a pseudo-Goldstone boson, the QCD axion mass is protected by the non-linearly realized PQ symmetry, which by definition is primarily broken by QCD instantons. Consequently, the QCD axion acquires a mass given by~\cite{GrillidiCortona:2015jxo} 
\begin{equation}
m_a = 5.7 {\rm \, meV} \left( \frac{10^{9} \GeV}{f_a} \right) \, , 
\end{equation}
where $f_a$ is the axion decay constant, which must lie well above the electroweak scale for phenomenological reasons, leading to a typical axion mass well below the eV scale~\cite{DiLuzio:2020wdo}. Additionally, the QCD axion is an excellent cold DM candidate over large regions of parameter space, when produced in the early universe via the misalignment mechanism~\cite{Preskill:1982cy, Abbott:1982af, Dine:1982ah}.

The most general couplings of the axion to the SM fermions can be written as 
\begin{equation}
\label{axion}
{\cal L}_{a}  =  \frac{\partial_\mu a}{2 f_a} \, \overline{f}_i \gamma^\mu \left( C^V_{ij} + C^A_{ij} \gamma_5 \right)f_j \, ,
\end{equation}
where $C^{V,A}_{ij}$ are hermitian matrices in flavor space. This Lagrangian can be mapped to the couplings in Eq.~\ref{scalar} by fermion field redefinitions, which apart from anomalous couplings to the SM gauge bosons, give rise to the identifications
\begin{equation}
\frac{\CC{\rS }{ij}}{\Lambda}  =  \frac{C^V_{ij}}{2 f_a}  \, , \qquad \qquad \qquad \frac{\CC{\rS5 }{ij}}{\Lambda}  =  \frac{C^A_{ij}}{2 f_a} \, .
\end{equation}
 The flavor-violationg couplings $C^{V,A}_{i \ne j}$ are determined by rotating the flavor-diagonal PQ charge matrices into the fermion mass basis (see Ref.~\cite{MartinCamalich:2020dfe} for details). In common QCD axion benchmark models, flavor alignment is realized either because the PQ charges of the SM fermions  vanish (``KSVZ models"~\cite{Kim:1979if,Shifman:1979if}) or PQ charges are taken to be flavor-universal  (standard ``DFSZ models"~\cite{Dine:1981rt,Zhitnitsky:1980tq}). However, the PQ charges in the DFSZ models do not have to be flavor-universal, and may constitute a new source of flavor violation beyond the SM Yukawas~\cite{Bardeen:1986yb,Geng:1988nc,Hindmarsh:1997ac}. In such scenarios, the size of flavor-violating axion couplings depends on the magnitude of the flavor misalignments parametrized by the unitary matrices that diagonalize the SM Yukawas. Thus, in the absence of a theory of flavor, these rotations are just described by a variety of new free parameters, which can be chosen suitably to realize an arbitrary pattern of flavor structures $C^{V,A}_{ij}$. Particular flavor patterns have been employed to e.g.~suppress the axion couplings to nucleons~\cite{DiLuzio:2017ogq,Bjorkeroth:2019jtx} and address stellar cooling anomalies~\cite{Saikawa:2019lng}, possibly correlated with low-energy signals from the extra DFSZ Higgs doublet~\cite{Badziak:2021apn}.     

Particularly motivated and predictive scenarios emerge when the PQ symmetry is identified with a flavor symmetry that addresses the SM flavor puzzle~\cite{Davidson:1981zd, Wilczek:1982rv, Berezhiani:1989fp, Ema:2016ops, Calibbi:2016hwq}. For instance, in the simplest realization, the PQ symmetry is identified with a $U(1)_F$ Froggatt-Nielsen symmetry~\cite{Froggatt:1978nt}, which necessarily possesses a QCD anomaly~\cite{Ibanez:1994ig,Binetruy:1994ru}. This setup allows one to predict flavor-violating axion couplings up to model-dependent ${\cal O}(1)$ coefficients~\cite{Ema:2016ops,Calibbi:2016hwq}. In particular, it determines the most phenomenologically relevant coupling, $C^V_{sd}$ (cf.~Section~\ref{CL}), to be on the order of the Cabibbo angle, $C^V_{sd} \sim V_{us} \sim 0.2$. Stronger suppression of light quark transitions occurs in models with non-abelian flavor symmetries; for example, in $U(2)$ models, $C^V_{sd} \sim V_{td} V_{ts} \sim 10^{-4}$~\cite{Linster:2018avp}. For the lepton sector several models have been presented in Ref.~\cite{Calibbi:2020jvd}, which can naturally yield large lepton flavor violation (LFV), for example  with couplings $C^V_{\mu e} \sim V_{us} \sim 0.2$ or $C^V_{\mu e} \sim \sqrt{m_e/m_\mu} \sim 0.1$. 

Note that even in scenarios with flavor alignment, renormalization group running induces flavor violation proportional to the CKM angles~\cite{Choi:2017gpf,MartinCamalich:2020dfe,Chala:2020wvs,Bauer:2021mvw}. This leads to strongly suppressed off-diagonal couplings, e.g. $C^V_{sd} \sim y_t^2/(16 \pi^2) V_{td} V_{ts} C_{tt}^V \log {\Lambda_{\rm UV}/\Lambda_{\rm IR}} \sim 10^{-5}$, which are phenomenologically irrelevant for axions masses $m_a \ll \keV$, since astrophysical constraints on the flavor-diagonal couplings yield significantly stronger constraints - see Sec.~\ref{AC}.

If there are substantial sources of explicit PQ breaking beyond QCD instantons, typically the axion acquires a large mass and does not solve the strong CP problem\footnote{Unless one employs further model-building, with e.g. new gauge sectors  confining at large energies with aligned vacuum angles, see Ref.~\cite{DiLuzio:2020wdo} for an overview of such models.}, and is called an axion-like partice (ALP). Nevertheless the ALP can still be a good DM candidate, provided its lifetime is sufficiently long, which can be produced  in the early universe by e.g. misalignment,   thermal freeze-in~\cite{Hall:2009bx} or parametric resonance~\cite{Co:2017mop}. In general, such ALPs can also have the same flavor-violating couplings in Eq.~\ref{axion} as the QCD axion, and thus are constrained in just the same way, provided their mass is below the experimental resolution. A flavor-violating ALP also has the same theoretical motivation as the QCD axion, possibly being a pseudo-Goldstone boson associated with global flavor-symmetries that address the Yukawa hierarchies, such as the ``Familon"~\cite{Wilczek:1982rv}, ``Flaxion"~\cite{Ema:2016ops}, ``Axiflavon"~\cite{Calibbi:2016hwq} or ``Froggatt-Nielsen ALP"~\cite{Greljo:2024evt}. Such a light ALP could arise from the spontaneous breaking of other global symmetries like baryon number~\cite{Heeck:2020nbq} or, more prominently, lepton number, in which case it is called the ``Majoron"~\cite{Chikashige:1980ui, Schechter:1981cv}. In this scenario large LFV couplings can be connected to the origin of neutrino masses in low-energy seesaw models - see Ref.~\cite{Calibbi:2020jvd} for an example. A light scalar with flavor-violating couplings could also be motivated as one of the moduli predicted by superstring compactifications~\cite{Cicoli:2013ana}, possibly related to modular flavor symmetries~\cite{Feruglio:2024dnc},  which have been employed for the flavor puzzle~\cite{Feruglio:2017spp} or to solve the strong CP problem with spontaneous CP violation~\cite{Feruglio:2023uof}.  

\subsection{Light Vector DM}

A light vector particle is also a viable DM candidate~\cite{Nelson:2011sf, Arias:2012az} which can be produced in the early universe by e.g. misalignment, thermal freeze-in or parametric resonance, as in the case of the ALP. These particles are often referred to in the literature as ``Dark Photon", ``Hidden Photon" or ``Dark $Z^\prime$". A dark photon usually refers to a vector that inherits all couplings to SM fermions from mixing with the photon and thus couples mainly to diagonal flavors. Instead general light vectors couple to fermions when these carry non-trivial charges under the dark gauge group,  which  generically induces flavor-violating couplings of the form in Eq.~\ref{vector} when these charges are not universal (see Ref.~\cite{Eguren:2024oov} for more details). This scenario is particularly motivated when the dark gauge group is identified with an anomaly-free flavor symmetry group shaping the structure of Yukawa matrices (see, e.g. Refs.~\cite{Smolkovic:2019jow, Bonnefoy:2019lsn, Greljo:2023bix}) or flavor non-universal charges such as $U(1)_{L_i - L_j}$~\cite{Foot:1994vd, Ardu:2022zom}. While the breaking scale of the gauge group has to be typically much larger than the electroweak scale (for a counter-example see Ref.~\cite{Grinstein:2010ve}), the associated massive gauge boson can be  light enough for our purposes if the gauge coupling is sufficiently small. Instead dipole couplings of the form in Eq.~\ref{dipole} can arise in models where the dark photon is extremely light, so that kinetic mixing is suppressed and the dominant couplings to SM fermions is through  higher-dimensional operators, see e.g. Ref.~\cite{Fabbrichesi:2020wbt}. 

\subsection{Axion and Vector Portals}

The flavor phenomenology  of the couplings in Eq.~\ref{setup} cannot distinguish between bosons that are dark and bosons that promptly decay into  dark particles. These could be part of a larger theoretical structure  (the dark sector), to which the bosons in Eq.~\ref{setup} would merely act as mediators, or  ``portals"~\cite{Patt:2006fw, Batell:2009di, Lanfranchi:2020crw}. An appealing theoretical feature here is that these lighter dark particles can easily be cosmologically stable as a result of some conserved quantum number in the dark sector, such as  dark fermion number. A new  phenomenological aspect  is that the (typically long-lived) bosonic mediator could also decay into SM particles.
The light mediator can then be created on-shell from decays of e.g. $B$- or $D$-mesons (copiously produced at hadron colliders), and may travel macroscopic distances before decaying visibly within e.g. the SHiP detector~\cite{Alekhin:2015byh, SHiP:2015vad}   located at the SPS or dedicated forward detectors at the LHC such as FASER~\cite{Feng:2017uoz, FASER:2018bac}.  This gives rise to a rich phenomenology that  correlates  signals at flavor factories like NA62 and Belle II with  essentially background-free searches at next-generation beam-dump experiments~\cite{Dobrich:2018jyi, Balkin:2024qtf}.


\section{CURRENT LIMITS} 
\label{CL}

The couplings in Eq.~\ref{setup} give rise to two-body decays of SM particles, such as $K \to \pi X$ or $\mu \to e X$, where $X$ denotes a dark scalar or dark vector boson, which would manifest as missing energy in a laboratory experiment. Thus, the experimental signature resembles SM decays involving a neutrino pair, but with a monochromatic visible particle. Its energy is essentially determined   by the mass of the invisible particle, which we assume to be approximately massless (i.e. with a mass below the experimental resolution).

The flavor-violating decay rate of a particle with mass $m$ into light bosons scales as $m^3/\Lambda_{\rm UV}^2$ because it arises from dimension-five operators. On the other hand, a flavor-changing SM decay rate scales as $\propto m^5 G_F^2$, with possible contributions from new heavy particles  scaling as $\propto m^5/\Lambda_{\rm UV}^4$, because both arise from dimension-six operators. Furthermore, SM amplitudes can be reduced by additional small factors due to chirality, loop effects, phase space, and/or CKM matrix elements, leading to a significant suppression relative to the two-body decay rate. Consequently, two-body missing energy searches can have an enormous sensitivity to the flavor-violating couplings in Eq.~\ref{setup}.

For example, the ratio of muon decay rates is given by
\begin{equation}
\frac{\Gamma (\mu \to e X)}{\Gamma (\mu \to e \nu \overline{\nu})} \approx \frac{m_\mu^3  (|\CC{\rX }{\mu e}|^2 + |\CC{\rX5 }{\mu e}|^2)/(16 \pi \Lambda^2)}{m_\mu^5G_F^2/(192 \pi^3)} = \left(\frac{9 \times 10^6 \GeV}{\Lambda_{\mu e}^{\rX}}  \right)^2 \, , 
\end{equation} 
where we neglected the electron mass, defined $\Lambda_{\mu e}^\rX=\Lambda/\sqrt{|\CC{\rX }{\mu e}|^2 + |\CC{\rX5 }{\mu e}|^2}$ and $X$ denotes either a massless dark scalar (for which $\CC{\rX, \rX5 }{\mu e} = \CC{\rS, \rS5 }{\mu e} $) or a massless dark vector (for which $\CC{\rX , \rX5}{\mu e} = \CC{\rV , \rV5}{\mu e}$). This means that the muon lifetime alone already sets a bound on the UV scale of order $10^7$ GeV. 

\begin{figure}[h]
\includegraphics[width=5in]{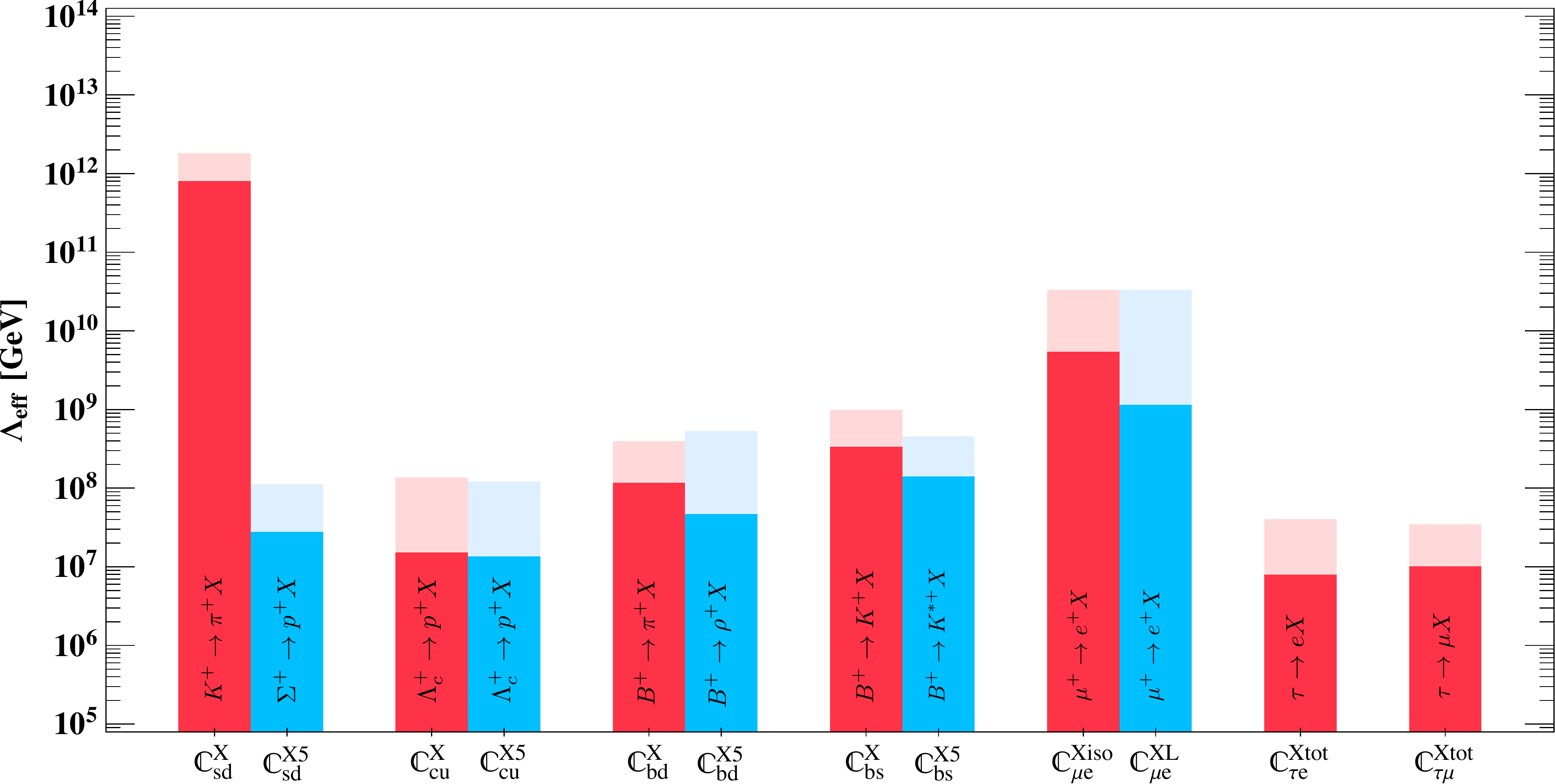}
\caption{Present limits on scalar and vector couplings for a massless dark boson. For a given quark flavor transition $i \to j$ and chirality structure X,X5 the bound is $\Lambda_{\rm eff} =  \Lambda/|\CC{\rS, \rS5}{ij}|$ for massless scalars  (cf.~Eq.~\ref{scalar}), $\Lambda_{\rm eff} =  \Lambda/|\CC{\rV, \rV5}{ij}|$ for massless vectors  (cf.~Eq.~\ref{vector}) and $\Lambda_{\rm eff} =  2 f_a /|C_{ij}^{V,A}|$ for derivative axion couplings (cf.~Eq.~\ref{axion}). For lepton transitions $\Lambda_{\rm eff} =  \Lambda/|\CC{\rX a}{ij}|$ with $\CC{\rX {\rm tot}}{ij} = \sqrt{|\CC{\rX}{ij}|^2 + |\CC{\rX5}{ij}|^2}$, $\CC{\rX {\rm iso}}{ij} = \CC{\rX}{ij}$ or $\CC{\rX5}{ij}$, $\CC{\rX {\rm L}}{ij} = \CC{\rX {\rm tot}}{ij}|_{\CC{\rX}{ij} = - \CC{\rX5}{ij}} $ for massless scalars  and vectors and similar for derivative couplings.}
\label{money1}
\end{figure}

\begin{figure}[h]
\includegraphics[width=5in]{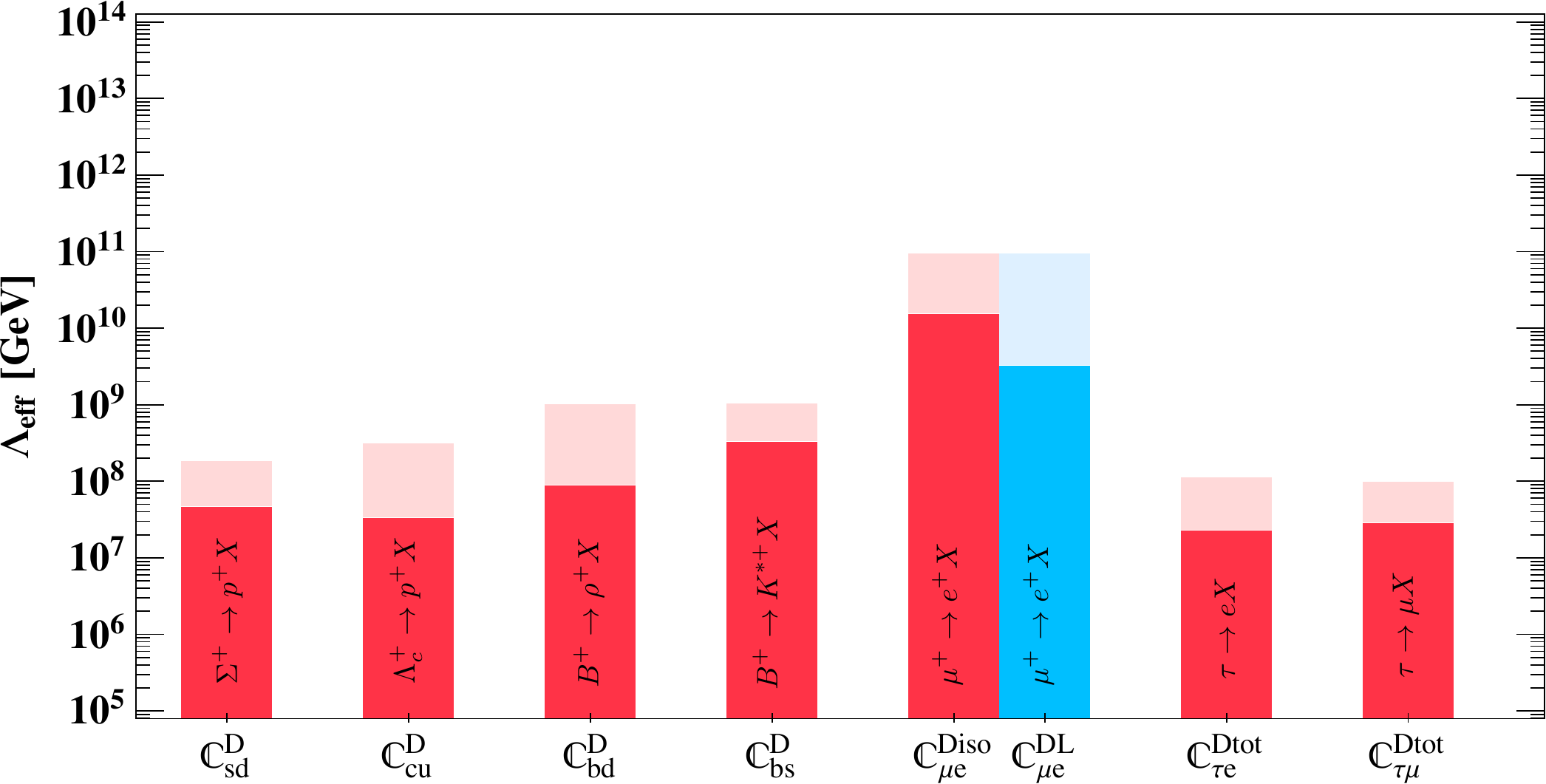}
\caption{Present limits on dipole couplings for a massless dark vector. For a given quark flavor transition $i \to j$ the bound is independent on the chiral structure $\Lambda_{\rm eff} =  \Lambda/|\CC{\rD, \rD5}{ij}|$. For lepton transitions  $\Lambda_{\rm eff} =  \Lambda/|\CC{\rD a}{ij}|$ with $\CC{\rD {\rm iso}}{ij} = \CC{\rD}{ij}$ or $\CC{\rD5}{ij}$, $\CC{\rD {\rm tot}}{ij} = \sqrt{|\CC{\rD}{ij}|^2 + |\CC{\rD5}{ij}|^2}$, $\CC{\rD {\rm L}}{ij} = \CC{\rD {\rm tot}}{ij}|_{\CC{\rD}{ij} = - \CC{\rD5}{ij}} $. }
\label{money2}
\end{figure}

Dedicated searches for two-body decays with missing energy give limits that are even more stringent, and are displayed in Figs.~\ref{money1} and \ref{money2}, where we collect the best current limits\footnote{For hadron decays we neglect the systematic errors on the associated form factors for simplicity which are, nonetheless, small for the decays of interest and have no major impact in the bounds.} on the couplings to massless dark bosons from laboratory experiments for each flavor transition, see also Tables~\ref{tableSV} and \ref{tableD}. We report and update previous limits obtained in Ref.~\cite{Feng:1997tn, Bjorkeroth:2018dzu,MartinCamalich:2020dfe, Calibbi:2020jvd, Carmona:2021seb, Bauer:2021mvw} for dark scalars, Refs.~\cite{Eguren:2024oov, Ibarra:2021xyk, Heeck:2016xkh} for dark vectors and Refs.~\cite{Gabrielli:2016cut, Fabbrichesi:2017vma, Su:2019ipw, Fabbrichesi:2020wbt, Su:2020xwt, Camalich:2020wac} for massless dark vectors with dipole couplings.

In the following, we discuss the experimental origin of these bounds in more detail, considering separately the couplings to quarks and to leptons. This separation is useful because in the quark sector the strongest limits typically come from pseudoscalar meson decays, which are sensitive to the chiral structure in Eq.~\ref{setup}, while in the lepton sector total decay rates do not depend on chirality. However, at least in the case of muons, one can rely on polarization, which gives control on the different couplings via the angular distribution of the final state lepton. In contrast to the quark sector the SM background is huge, and thus it is convenient to constrain the couplings depending on whether the chiral structure is aligned to the SM  (i.e. left-handed couplings with ${\CC{\rX}{ij} = - \CC{\rX5}{ij}} $) or not, and we restrict for simplicity to isotropic decays, corresponding to couplings with either $\CC{\rX}{ij} = 0$ or $\CC{\rX5}{ij}  = 0$.

\begin{table}[h]
\renewcommand{\arraystretch}{1.4}
  \setlength{\arrayrulewidth}{.25mm}
\centering
\setlength{\tabcolsep}{1 em}
\caption{Present and expected laboratory  limits on effective couplings of massless dark scalars and vectors in units of GeV, with the same notation as in Fig.~\ref{money1}. See text for details. \label{tableSV}}
\vspace{0.5cm}
\label{limitsSV}
\begin{tabular}{@{}|c|c|c|c|c|c|c|c|c|c|c|c|c|@{}}
\hline
& $\CC{\rX}{\rm sd}$ &  $\CC{\rX 5}{\rm sd}$  & $\CC{\rX}{\rm cu}$ &  $\CC{\rX 5}{\rm cu}$  \\
\hline
$\Lambda_{\rm eff}$ & $8.1 \times 10^{11}$~\cite{NA62:2021zjw} &  $2.8 \times 10^{7}$~\cite{BESIII:2023utd} & $1.5 \times 10^{7}$~\cite{BESIII:2022vrr} & $1.4 \times 10^{7}$~\cite{BESIII:2022vrr}   \\
$\Lambda_{\rm eff}^{\rm proj}$ & $1.8 \times 10^{12}$~[NA62] & $1.1 \times 10^{8}$~[STCF]  & $1.4 \times 10^{8}$~[STCF] & $1.2 \times 10^{8}$~[STCF]  \\
\hline
\hline
& $\CC{\rX}{\rm bd}$ &  $\CC{\rX 5}{\rm bd}$ & $\CC{\rX}{\rm bs}$ &  $\CC{\rX 5}{\rm bs}$  \\
\hline
$\Lambda_{\rm eff}$ & $1.2 \times 10^{8}$~\cite{BaBar:2004xlo, MartinCamalich:2020dfe} & $4.7 \times 10^{7}$~\cite{ALEPH:2000vvi, Alonso-Alvarez:2023mgc} & $3.4 \times 10^{8}$~\cite{Belle-II:2023esi, BaBar:2013npw, Altmannshofer:2023hkn} & $1.4 \times 10^{8}$~\cite{BaBar:2013npw, Altmannshofer:2023hkn}  \\
$\Lambda_{\rm eff}^{\rm proj}$ & $4.0 \times 10^{8}$~[Belle II] & $5.4 \times 10^{8}$~[Belle II] & $1.0 \times 10^{9}$~[Belle II] & $4.6 \times 10^{8}$~[Belle II] \\
\hline
\hline
 &  $\CC{\rX \rm iso}{\rm \mu e }$ &  $\CC{\rX \rm L}{\rm \mu e}$ &  $\CC{\rX \rm tot}{\rm \tau e}$ &  $\CC{\rX \rm tot}{\rm \tau \mu}$ \\
 \hline
$\Lambda_{\rm eff}$ & $5.5 \times 10^{9}$~\cite{Jodidio:1986mz} & $1.2 \times 10^{9}$~\cite{Bayes:2014lxz}   & $8.0 \times 10^{6}$~\cite{Belle-II:2022heu}  & $1.0 \times 10^{7}$~\cite{Belle-II:2022heu}  \\
$\Lambda_{\rm eff}^{\rm proj}$   & $3.3 \times 10^{10}$~[Mu3e] & $3.3 \times 10^{10}$~[Mu3e] & $4.0 \times 10^{7}$~[Belle II] & $3.5 \times 10^{7}$~[Belle II] \\
\hline
\end{tabular}
\end{table}
\begin{table}[h]
\renewcommand{\arraystretch}{1.4}
  \setlength{\arrayrulewidth}{.25mm}
\centering
\setlength{\tabcolsep}{1 em}
\caption{Present and expected laboratory  limits on effective dipole couplings of massless dark vectors in units of GeV, with the same notation as in Fig.~\ref{money2}. See text for details. \label{tableD}}
\vspace{0.5cm}
\label{limitsDP}
\begin{tabular}{@{}|c|c|c|c|c|c|c|c|c|c|c|c|c|@{}}
\hline
& $\CC{\rD}{\rm sd}$   & $\CC{\rD}{\rm cu}$ &   $\CC{\rD}{\rm bd}$ &  $\CC{\rD}{\rm bs}$ \\
\hline
$\Lambda_{\rm eff}$ & $4.7 \times 10^{7}$~\cite{BESIII:2023utd} &  $3.3 \times 10^{7}$I~\cite{BESIII:2022vrr} & $9.0 \times 10^{7}$~\cite{ALEPH:2000vvi, Alonso-Alvarez:2023mgc}  & $3.3 \times 10^{8}$~\cite{BaBar:2013npw,Altmannshofer:2023hkn} \\
$\Lambda_{\rm eff}^{\rm proj}$ & $1.9 \times 10^{8}$~[STCF] & $3.2 \times 10^{8}$~[STCF] & $1.0 \times 10^{9}$~[Belle II] & $1.1 \times 10^{9}$~[Belle II] \\
\hline
\hline
&   $\CC{\rD \rm iso}{\rm \mu e }$ &  $\CC{\rD \rm L}{\rm \mu e}$ &  $\CC{\rD \rm tot}{\rm \tau e}$ &  $\CC{\rD \rm tot}{\rm \tau \mu}$\\
\hline
$\Lambda_{\rm eff}$  & $1.6 \times 10^{10}$~\cite{Jodidio:1986mz}  & $3.3 \times 10^{9}$~\cite{Bayes:2014lxz} & $2.3 \times 10^{7}$~\cite{Belle-II:2022heu} & $2.9 \times 10^{7}$~\cite{Belle-II:2022heu} \\
$\Lambda_{\rm eff}^{\rm proj}$ & $9.5 \times 10^{10}$~[Mu3e] & $9.5 \times 10^{10}$~[Mu3e] & $1.1 \times 10^{8}$~[Belle II] & $9.9 \times 10^{7}$~[Belle II]  \\
\hline
\end{tabular}
\end{table}
\subsection{Quark Sector}
\label{CL:quarks}

In the quark sector the strongest constraints typically arise from laboratory searches for two-body decays of pseudoscalar mesons and baryons with an invisible scalar or vector particle in the final state. These decay rates scale with the couplings according to
\begin{equation}
\Gamma_{P\to P^{\prime} X} \propto |\CC{\rX}{ij} |^2\, , \qquad
\Gamma_{P\to V X} \propto  |\CC{\rX 5}{ij} |^2 \, , \qquad
\Gamma_{B \to B^{\prime}X} \propto f_1^2 |\CC{\rX}{ij} |^2 +  g_1^2 |\CC{\rX5}{ij} |^2\, , 
\end{equation}
where $P$ and $P^\prime$ denote pseudo-scalar mesons, $V$ is a vector meson,  $B, B^\prime$ are baryons, $f_1$ and $g_1$ are baryonic form factors, and we use the same notation for the couplings as above. For the complete expressions and a  collection of the relevant form factors see Refs.~\cite{Eguren:2024oov, MartinCamalich:2020dfe}. 

In order to infer the limits in Fig.~\ref{money1} and \ref{money2}, these rates are computed for a single coupling switched on at a time and compared to the experimental limits on the various decays summarized in Table~\ref{exp}. Often, the experimental collaborations do not provide limits on two-body decays with missing energy; nevertheless, in some cases there is enough information to extract this bound from available data. We indicate the limits obtained in this way by a checkmark in the last column, where we used the results from the indicated references - see Refs.~\cite{Eguren:2024oov, MartinCamalich:2020dfe} for more details. It would be of course preferable if these limits were replaced by proper experimental analyses. 

\begin{table}[h]
\renewcommand{\arraystretch}{1.5}
  \setlength{\arrayrulewidth}{.25mm}
\centering

\setlength{\tabcolsep}{1 em}
\caption{Laboratory limits on the branching ratios of two-body meson, baryon and lepton decays with $X$ denoting a massless invisible particle (i.e. any mass below the mass resolution of the experiments). For  polarized muon decays the limit depends on the angular distribution of the electron, denoted as ${\rm iso}$ for isotropic decays, $R$ for $\propto 1 +  \cos \theta$ and $L$ for $\propto 1 -  \cos \theta$ (as in the SM). }
\label{exp}
\begin{center}
\begin{tabular}{@{}|c|c|c|l|@{}}
\hline
Decay & 90\% CL Limit & Reference  & Recast \\
\hline
${\rm BR} (K \to\pi X)$ &  $5.0 \times 10^{-11}$ &NA62~\cite{NA62:2021zjw} & {\color{green} $\times$}   \\
${\rm BR} (\Sigma  \to p X)$ & $3.2 \times 10^{-5}$ & BESIII~\cite{BESIII:2023utd}$^{\rm a}$ & {\color{green} $\times$} \\
${\rm BR} (D  \to \pi X)$ & $8.0 \times 10^{-6}$& CLEO~\cite{CLEO:2008ffk}  & {\color{red} $\checkmark$}~\cite{MartinCamalich:2020dfe}  \\
${\rm BR} (\Lambda_c  \to p X)$ & $8.0 \times 10^{-5}$& BESIII~\cite{BESIII:2022vrr}  & {\color{green} $\times$}  \\
${\rm BR} (B  \to K X)$ &$7.0 \times 10^{-6}$ & Belle~II~\cite{Belle-II:2023esi}  \& BaBar ~\cite{BaBar:2013npw}  & {\color{red} $\checkmark$}~\cite{Altmannshofer:2023hkn} \\
${\rm BR} (B  \to K^* X)$ &$4.2 \times 10^{-5}$ & BaBar~\cite{BaBar:2013npw}   & {\color{red} $\checkmark$}~\cite{Altmannshofer:2023hkn} \\
${\rm BR} (B  \to \pi X)$ & $2.3 \times 10^{-5}$& BaBar~\cite{BaBar:2004xlo} & {\color{red} $\checkmark$}~\cite{MartinCamalich:2020dfe}  \\
${\rm BR} (B  \to \rho X)$ & $3.9 \times 10^{-4}$& LEP~\cite{ALEPH:2000vvi}     & {\color{red} $\checkmark$}~\cite{Alonso-Alvarez:2023mgc} \\
\hline
${\rm BR} (\mu \to e X)_{\rm iso}$ &  $2.6 \times 10^{-6}$ & TRIUMF~\cite{Jodidio:1986mz} &   {\color{green} $\times$} \\
${\rm BR} (\mu \to e X)_R$ &  $2.5 \times 10^{-6}$ & TRIUMF~\cite{Jodidio:1986mz} &   {\color{red} $\checkmark$}~\cite{Calibbi:2020jvd}  \\
${\rm BR} (\mu \to e X)_L$ &  $5.8 \times 10^{-5}$ &TWIST~\cite{Bayes:2014lxz}  & {\color{green} $\times$}   \\
${\rm BR} (\tau  \to \mu X)$ & $4.7 \times 10^{-4}$ & Belle II~\cite{Belle-II:2022heu} &  {\color{green} $\times$}  \\
${\rm BR} (\tau  \to e X)$ & $7.6 \times 10^{-4}$ & Belle II~\cite{Belle-II:2022heu} &  {\color{green} $\times$}  \\
\hline
\end{tabular}
\end{center}
\begin{tabnote}
$^{\rm a}$ SN~1987A cooling constrains hyperon decays, ${\rm BR} (\Lambda  \to n X) \lesssim 8 \times 10^{-9}$~\cite{Camalich:2020wac} - see Section~\ref{AC}; 
\end{tabnote}
\end{table}

For $K \to \pi X$ and $\Sigma \to p X$ we use the experimental limits for invisible massless $X$ reported by NA62~\cite{NA62:2021zjw} and 
BESIII~\cite{BESIII:2023utd}, respectively. Note that for the former decay a slightly stronger limit ${\rm BR}(K^+\to\pi^+X)<2.8\times10^{-11}$ (at 90\% C.L.) was recently derived in Ref.~\cite{Guadagnoli:2025xnt} by recasting the publicly available NA62 dataset collected between 2016 and 2022. This leads to a constraint on the effective coupling that is tighter than the one shown in Tab.~\ref{tableSV}, $\CC{\rX}{\rm sd}>1.1\times 10^{12}$ GeV.

The recast of the $D\to\pi \nu\bar\nu$ decays require a more detailed discussion. In Ref.~\cite{MartinCamalich:2020dfe} a bound on the two-body decay was derived from a null search performed by CLEO of the decay $D^+\to\tau^+(\to\pi^+\bar\nu)\nu$~\cite{CLEO:2008ffk} by recasting the bin in the pion spectrum corresponding to $m_X\approx0$. Since then, two analyses by the BESIII collaboration~\cite{BESIII:2019vhn,BESIII:2024vlt} have reported observations of the $D^+ \to \tau^+(\to \pi^+ \bar{\nu}) \nu$ decay mode. From the pion spectrum of the signal reported in these analyses, it is clear that a search for the two-body $D^+\to\pi^+ X$ decay at BESIII (superseding the current CLEO recast) requires a careful treatment of the SM background from tau decays.   

The BESIII collaboration has also reported a search of the three-body decay $D^0\to\pi^0\nu\bar\nu$ using 2.93 fb$^{-1}$ of data (roughly $1/7$ of the total luminosity planned to be collected at BESIII~\cite{BESIII:2020nme}) and obtained the upper limit $2\times10^{-4}$ at 90\% CL for the branching fraction. If one takes this as representative of the sensitivity achievable for ${\rm BR}(D^0 \to \pi^0 X)$, the resulting limit would be considerably weaker than the recast of the CLEO data for the charged mode. This highlights the importance of dedicated analyses of  $D \to \pi X$ decays in both charged modes for a robust search for dark bosons in charmed meson decays.

Finally, in the same context of probing  $cu$ couplings, the BESIII collaboration has performed a search of the baryonic two-body $\Lambda_c^+\to p X$ decays with their complete data set~\cite{BESIII:2022vrr}. This search becomes even more relevant in light of the limitations discussed above regarding current analyses of $D$-meson decays, and sets the best present limit on $cu$ couplings.

In case of the $bs$ and $bd$ couplings, there are no analyses to date of the invisible two-body decays of $B_{(s)}$ mesons or bottom baryons. Therefore, all the existing limits are obtained from recasts of searches of the three-body decays into neutrinos~\cite{MartinCamalich:2020dfe,Altmannshofer:2023hkn,Alonso-Alvarez:2023mgc}. The upper limits on $B\to K X$, $B\to K^*X$ and $B\to \pi X$ use BaBar data~\cite{BaBar:2013npw,BaBar:2004xlo}, while the upper limit on $B\to \rho X$ uses data from the ALEPH experiment at LEP~\cite{,ALEPH:2000vvi}. Only in case of $B^+\to K^+ X$ the recast uses also Belle~II data~\cite{Belle-II:2023esi}.

Three-body decays typically give weaker constraints than two body decays. For example, LHCb constraints on $B_{(s)} \to \mu \mu X$ cannot compete with Belle~II limits from $B\to K^{(*)} X$ or $B\to \pi X$ and  $B\to \rho X$ decays~\cite{Albrecht:2019zul}, while multi-hadron final states such as $B \to K \pi X$ are subject to larger theoretical uncertainties apart from experimental challenges. One exception is the kaon sector where the corresponding $P\to VX$ decay mode is kinematically inaccessible ($m_{\rho}>m_K$) and an important bound stems from $K\to\pi\pi X$ decays. Experimentally,  both $K^+\to\pi^0\pi^+ X$~\cite{Adler:2000ic,Sadovsky:2023cxu} and $K_L\to\pi^0\pi^0 X$~\cite{E391a:2011aa} have been searched for. Theoretically, the decay rate into scalars can be predicted robustly using isospin symmetry and the form factors determined from the measurements of $K\to\pi\pi e\nu$ charge-current decays~\cite{MartinCamalich:2020dfe} (see also Refs.~\cite{Batley:2010zza,Batley:2012rf,Littenberg:1995zy,Chiang:2000bg,Cavan-Piton:2024pqp}). In fact, the strongest laboratory constraint on $\CC{\rS5}{sd}$ stems from the OKA upper limit on  $K^+\to\pi^0\pi^+ X$~\cite{Sadovsky:2023cxu} - see Tab.~\ref{exp}.~\footnote{The decay mode $K_L\to\pi^0\pi^0 X$ in Ref.~\cite{E391a:2011aa} could nominally lead to a stronger bound but the analysis does not cover the range $m_X\leq50$ MeV and cannot be used to derive a bound for the massless $X$ case.} In principle, a similar analysis can be performed for decays involving a dark vector with dipole couplings. However, this scenario requires the computation of form factors in QCD, whereas current results rely only on approximate estimates derived from quark models
~\cite{Fabbrichesi:2017vma}.   

Finally, neutral meson mixing, which gives the most stringent constraints on dimension-six SMEFT operators induced by heavy new physics~\cite{Bona:2024bue}, yields limits on flavor-violating couplings that are at most  comparable to those obtained from two-body decays, but are subject to uncertainties due to contributions from UV physics that are parametrically of the same order~\cite{MartinCamalich:2020dfe}. For example, the radial mode in ALP models contributes to mixing amplitudes parametricallly at the same level as the ALP itself, since its mass is set by the same UV scale that suppresses the couplings of the tree-level ALP exchange.

\subsection{Lepton Sector}

In the lepton sector, the strongest constraints arise from laboratory searches for two-body LFV decays. The decay rates scale with the  sum of the squared couplings
\begin{equation}
\Gamma_{\ell \to \ell^{\prime} X} \propto |\CC{\rX}{\ell^\prime \ell} |^2 +  |\CC{\rX5}{\ell^\prime \ell} |^2\,. 
\end{equation}
The experimental difficulty is that these decays look very similar to the corresponding SM decays, resulting in a single visible object plus missing energy. As a consequence, the $\ell\to \ell' X$ decays are not covered by standard LFV searches and require dedicated experimental strategies to suppress the large SM background. One possibility is to study decays of polarized leptons, which gives an angular distribution of the final state lepton that allows to distinguish between the chiral LFV couplings since~\cite{Calibbi:2020jvd, Eguren:2024oov}  
\begin{equation}
\label{poldecay}
\frac{d \Gamma (\ell \to \ell^\prime X)}{d \cos \theta}  \propto 1 +2 \cos\theta\cdot \frac{{\rm Re}( \CC{\rX }{\ell^\prime\ell} \CC{\rX5*}{\ell^\prime\ell})}{|\CC{\rX }{\ell^\prime\ell} |^2  + |\CC{\rX5}{\ell^\prime\ell}|^2} \, ,
\end{equation}
where $\theta$ is the angle between the polarization vector of the decaying lepton $\ell$ and the momentum of the final state lepton $\ell^\prime$. The three-body SM  decay rate for a final state lepton energy close to the maximal value $E_{\ell^\prime} = m_\ell/2$ is 
\begin{equation}
\frac{d \Gamma (\ell \to \ell^\prime \nu \overline{\nu})}{d \cos \theta}  \propto 1 -  \cos \theta \, , 
\end{equation}
due to the V-A structure of the SM. Therefore, at angles close to $\theta = \pi$ the SM background is strongly reduced, giving sensitivity to the two-body decay, unless it is aligned with the SM for $\CC{\rX5}{\ell^\prime\ell} = - \CC{\rX}{\ell^\prime\ell}$. The experimental collaborations that search for $\mu \to e X$ usually constrain only a given benchmark scenario, but using the complete data sets, one can constrain any combination of chiral couplings. Here we focus only on the limits of either isotropic decays  ($\CC{\rX}{\ell^\prime\ell} = 0$ or $\CC{\rX5}{\ell^\prime\ell} = 0$) or purely left-handed ($\CC{\rX5}{\ell^\prime\ell} = - \CC{\rX}{\ell^\prime\ell}$) couplings, and use the limits obtained in Ref.~\cite{Calibbi:2020jvd} from searches for massless invisible particles in muon decays carried out in the late 80's at TRIUMF.  These are severely weakened for couplings aligned with the SM decay and are replaced by the TWIST searches that rely on the monochromatic electron as the signal. Another handle is through three-body decays with an extra photon, $\mu \to e \gamma X$, which at present gives weaker constraints than the two-body decays, but show interesting prospects to increase the bound using MEG-II data, as discussed in Sec. \ref{FP}. We thus show also the present limit from the Crystal Ball collaboration~\cite{Bolton:1988af}. Finally, for tau lepton decays only limits on the total branching ratio have been obtained recently at Belle-II~\cite{Belle-II:2022heu}. We summarize the best current limits on LFV decays in Table~\ref{exp}, which are used together with the predictions for the decay rates in Ref.~\cite{Calibbi:2020jvd} and \cite{Eguren:2024oov} to derive the limits in Figs.~\ref{money1} and \ref{money2}, with a single coupling switched on at a time.


\section{FUTURE PROSPECTS}
\label{FP}

We now review the prospects of future searches for light particles in flavor-violating decays with missing energy. Because of the very different experimental challenges, we again split the discussion into quark and lepton sector. In general the sensitivity can be improved by: 1) performing existing searches with larger data sets; 2) applying dedicated search strategies to existing data sets; and, 3) performing entirely new searches.  

\subsection{Quark Sector}
\label{qfut}
\begin{table}[h]
\renewcommand{\arraystretch}{1.5}
  \setlength{\arrayrulewidth}{.25mm}
\centering
\setlength{\tabcolsep}{1 em}
\caption{Forecast of future laboratory  limits on branching ratios of two-body meson, baryon and lepton decays with $X$ denoting a massless invisible particle. }
\vspace{0.5cm}
\label{expfut}
\begin{tabular}{@{}|l|c|c|@{}}
\hline
Decay  & Prospective Limit & Experiment$^a$   \\
\hline
${\rm BR}_{\rm proj} (K \to\pi X)$ &  $1 \times 10^{-11}$ &NA62~\cite{MartinCamalich:2020dfe}   \\
${\rm BR}_{\rm proj} (K \to\pi\pi X)$ &  $7\times10^{-7}$ &E391a~\cite{E391a:2011aa,MartinCamalich:2020dfe}   \\
${\rm BR}_{\rm proj} (\Sigma  \to p X)$ &  $2 \times 10^{-6}$ & STCF \\
${\rm BR}_{\rm proj} (D  \to \pi X)$ & $1 \times  10^{-5}$ & STCF    \\
${\rm BR}_{\rm proj} (\Lambda_c  \to p X)$ &$1 \times  10^{-6}$ & STCF   \\
${\rm BR}_{\rm proj} (B  \to K X)$ & $8\times10^{-7}$& Belle II \\
${\rm BR}_{\rm proj} (B  \to K^* X)$ & $4\times10^{-6}$ &Belle II  \\
${\rm BR}_{\rm proj} (B  \to \pi X)$ &  $2\times10^{-6}$ &Belle II    \\
${\rm BR}_{\rm proj} (B  \to \rho X)$ & $3 \times10^{-6}$ & Belle II      \\
\hline
${\rm BR} (\mu  \to e X)_{\rm iso}$ &$7 \times 10^{-8}$ & MEGII-fwd~\cite{Calibbi:2020jvd}    \\
${\rm BR} (\mu  \to e X)_{{\rm iso}, L, R}$ &$7 \times 10^{-8}$ & Mu3e~\cite{Perrevoort:2018ttp, Perrevoort:2018okj}   \\
${\rm BR} (\mu  \to e X)_{\rm iso}$ &$1 \times  10^{-7}$ & Mu2e-X, COMET-X~\cite{Hill:2023dym}    \\
${\rm BR} (\tau  \to e X)$ & $4 \times 10^{-5}$ & Belle II~\cite{Calibbi:2020jvd, Badziak:2024szg}    \\
${\rm BR} (\tau  \to \mu X)$ & $3 \times 10^{-5}$ & Belle II~\cite{Calibbi:2020jvd, Badziak:2024szg}    \\
\hline
\end{tabular}
\begin{tabnote}
$^{\rm a}$ In brackets we indicate  the references where the forecasts were performed.
\end{tabnote}
\end{table}

Future sensitivities to flavor-violating two-body hadron decays with missing energy can be estimated by rescaling existing searches with the expected luminosity increase. Starting with $K^+\to \pi^+ X$ and massless $X$, an improvement by an order of magnitude compared to the BNL result, ${\rm BR}(K^+\to \pi^+ X)<7.3 \times 10^{-11}$~\cite{Adler:2008zza}, can be expected using the full data set of NA62~\cite{MartinCamalich:2020dfe}, and we use ${\rm BR_{proj}}(K^+\to\pi^+ X)=10^{-11}$ as a conservative projection. The same flavor transition is probed by KOTO looking for the neutral decay mode $K_L\to \pi^0 X$. The KOTO collaboration expects the sensitivity to be improved down to the $10^{-11}$ level~\cite{Goudzovski:2022vbt}, which would provide a sensitivity to the $sd$ couplings  similar to NA62. Also searches for the $K_L\to \pi^0\pi^0 X$ mode could be performed by KOTO, although no feasibility study has been carried out yet. Extending the existing upper limit of the E391a collaboration on this mode~\cite{E391a:2011aa} to the case of massless $X$ would lead to the strongest $K\to\pi\pi X$ limit on the $sd$ couplings~\cite{MartinCamalich:2020dfe}.    

Another important probe for these transitions is offered by hyperon decays, which already give the best current limit on axial transitions using searches conducted by BESIII for massless vectors. The present limit of
${\rm BR} (\Sigma  \to p X) < 3.2 \times 10^{-5}$ already utilizes the full data set that BESIII has reported on the $J/\Psi$ resonance and significant improvements beyond this limit will require a new measurement campaign. For example, a Super Tau Charm Factory (STCF)~\cite{Zhou:2021rgi} with an integrated luminosity of 1 ab$^{-1}$ could produce a sample of $3.4\times10^{12}$ $J/\Psi$'s ($\sim3400$ times more than in BESIII). Our projection in the $\Sigma\to pX$ mode is then obtained by rescaling the current bound by the square root of this increase. Future searches of other decay modes such as $\Lambda\to nX$ can potentially reach a similar sensitivity. 

Many of the current BESIII bounds on two-body decays probing the $cu$ couplings would also be improved by increasing data sets or dedicated searches at a future STCF. For instance, in the case of the meson $D \to \pi X$ decays the situation could easily improve by performing the two-body analysis over the existing $D^+ \to \tau^+(\to\pi^+\bar\nu) \nu$ and $D^0 \to \pi^0 \nu \overline{\nu}$ data sets~\cite{BESIII:2019vhn,BESIII:2024vlt,BESIII:2021slf}. Moreover, the latter analysis only employed 2.93 fb$^{-1}$ corresponding to a sample of $10^7$ $D^0\bar D^0$ pairs. Rescaling the current upper limit on the $D^0 \to \pi^0 \nu \overline{\nu}$ decay~\cite{BESIII:2021slf} by the square root of the ratio of luminosities or number of $D^0$ mesons, one can forecast the future bounds that could be obtained with the 20 fb$^{-1}$ to be collected by the end of the planned BESIII operation~\cite{BESIII:2020nme}, ${\rm BR}_{\rm proj}^{\rm BESIII}(D^0\to\pi^0 X)=8\times 10^{-5}$, or the $3.6\times 10^9$ $D^0\bar D^0$ pairs aimed to be collected at a STCF~\cite{Zhou:2021rgi}, ${\rm BR}_{\rm proj}^{\rm STCF}(D^0\to\pi^0 X)=10^{-5}$. 

We note that these estimates of the projections for direct searches might be over-conservative, as stronger bounds could stem from a dedicated two-body analysis. Therefore, a forecast of the charmed baryon decays, where this has been performed with the full BESIII data set, is more direct. Rescaling by the square root of the sample size $\sim10^5$ $\Lambda_c$'s reported in the BESIII sample~\cite{BESIII:2022vrr} with $5.6\times10^8$ charmed baryons planned in an SCTF~\cite{Zhou:2021rgi} we find ${\rm BR}(\Lambda_c\to p X)\lesssim10^{-6}$.  We use this forecast for the projections of the $cu$ couplings shown in Figs.~\ref{money1} and \ref{money2}. 

Turning to $bs$ and $bd$ couplings, the sensitivity to branching ratios for invisible decay modes of $B$-mesons will improve significantly with Belle~II, where the final integrated luminosity is anticipated to reach 50~ab$^{-1}$. This increase in luminosity should also enable several dedicated analyses of invisible two-body decays, superseding the recasts currently used to derive the limits shown in Table~\ref{exp}.

In order to estimate the future sensitivity for the $B \to K X$ transition, we use the recast performed in Ref.~\cite{Altmannshofer:2023hkn} of the Belle~II search for $B^+ \to K^+ \nu \bar{\nu}$~\cite{Belle-II:2023esi}, rescaling it by the square root of the ratio of the associated luminosity (362~fb$^{-1}$) and the prospective size of the data set (50~ab$^{-1}$). For the rest of the two-body decay modes, future limits can be estimated by rescaling in a similar way the recasts of the three-body decays reported by BaBar with the corresponding gain in data sample size at Belle~II as done in Ref.~\cite{MartinCamalich:2020dfe} (which is approximately factor 100 assuming similar reconstruction efficiencies at Belle~II and BaBar)~\footnote{Note that in Ref.~\cite{MartinCamalich:2020dfe} an optimistic rescaling linear in the ratio of luminosities was used, while here we employ a more conservative scaling with the square-root of luminosities.}.  In the case of $B\to\rho X$, where the only current bound was obtained by a recast of LEP data in Ref.~\cite{Alonso-Alvarez:2023mgc},  the future projection is estimated as in Ref.~\cite{MartinCamalich:2020dfe}, using the current Belle bound on the three-body decay modes $B\to\rho\nu\bar\nu$ (for $711$ fb$^{-1}$)~\cite{Belle:2017oht}, but again employing square-root luminosity scaling instead of linear scaling. 

Finally, we want to emphasize that there are several potentially interesting channels to search for a dark boson signal, as suggested in Ref.~\cite{MartinCamalich:2020dfe}.  For example, there are no measurements of mesonic decay modes in $c\to u$ transitions that are sensitive to the axial-vector coupling, i.e.,  there are no $D \to \pi \pi X$ or $D \to \rho X$ searches with $X$ being an invisible massless particle or di-neutrino state. One could also search for a signal in $D_s\to K X$, $D_s\to K^* X$ decays, all of which could be performed at Belle II, BESIII and at STCF. 

\subsection{Lepton Sector}

Probes of SM predictions for rare processes with charged leptons will improve substantially in the next decade. The muon beam experiments MEG II \cite{Baldini:2018nnn}, Mu3e \cite{Berger:2014vba,Blondel:2013ia}, COMET~\cite{Adamov:2018vin} and Mu2e~\cite{Bartoszek:2014mya} will collect unprecedented datasets using $\mathcal{O}(10^{15}-10^{17})$ muons each. Similarly,  Belle II and STCF are expected to collect roughly $5\times10^{10}$~\cite{Perez:2019cdy}  and $ 2 \times 10^{10}$~\cite{Achasov:2023gey} $\tau^+\tau^-$ pairs, respectively, exceeding the datasets at Belle and BaBar by more than an order of magnitude.

Starting with two-body $\tau$-decays, simple estimates for future sensitivities can be obtained by upscaling the present Belle II bound with the full expected data set, as done in Refs.~\cite{Calibbi:2020jvd, Badziak:2024szg}. Using the current expected bound for ${\rm BR} (\tau \to \ell X)$ provided in Ref.~\cite{Belle-II:2022heu} for 62.8 fb$^{-1}$, one may estimate that Belle II with 50 ab$^{-1}$ may set 90 \%CL limits on  flavor-violating $\tau$-decays given by  ${\rm BR}_{\rm proj} (\tau \to e X) = 4 \times 10^{-5}$ and ${\rm BR}_{\rm proj} (\tau \to \mu X) = 3 \times 10^{-5}$.  Novel analysis strategies have been proposed to further improve signal/background discrimination by using suitable kinematic variables, which might strengthen these bounds by a factor of three~\cite{Guadagnoli:2021fcj} or even an order of magnitude~\cite{DeLaCruz-Burelo:2020ozf}.

Many new ideas have been put forward to increase sensitivity of searches for two-body muon decays at various muon beam experiments, see Ref.~\cite{Knapen:2023zgi} for a recent proposal for Mu3e searches for $\mu \to eee X$ and a concise overview of other recent ideas. Up to now the only experimental study~\cite{Perrevoort:2018ttp, Perrevoort:2018okj}
relies on an online trigger proposal optimized to look for monochromatic $\mu \to e X$ events on top of the three-body SM Michel spectrum at the Mu3e experiment, with an expected limit of ${\rm BR}_{\rm proj} (\mu \to e X)_{\rm iso} = 7 \times 10^{-8}$. This strategy  requires an extremely accurate control of theoretical uncertainties due to  the irreducible $\mu \to e \overline{\nu} \nu$ background, which has been quantified in Ref.~\cite{Banerjee:2022nbr} and implemented in the Monte
Carlo code McMule~\cite{Banerjee:2020rww}. 

While the Mu3e approach does not rely on polarization to suppress background, and thus is  independent on the specific chiral structure of dark boson couplings, it faces severe challenges related to systematics uncertainties in searching for a bump close to the endpoint of  the SM spectrum (corresponding to massless $X$), as this region is typically assumed to be signal-free and used for experimental calibration. For this reasons alternative calibration techniques and/or search strategies are required for Mu3e. One interesting proposal in this direction is to look for  $\mu \to eee X$ decays, which can be expected to be sensitive to flavor-violating $\mu e$ couplings at the same order as the current Mu3e proposal, but without the same experimental challenges.

Another proposal for a new experimental set-up at MEG II is MEGII-fwd, 
which consists of a dedicated calorimeter installed in the forward direction relative to the muon beamline~\cite{Calibbi:2020jvd}. The search strategy follows the experiment  by Jodidio et al. in 1986~\cite{Jodidio:1986mz}, looking for $\mu^+\to e^+a$ decays, utilizing that $\mu^+$ is polarized antiparallel to the beam direction, up to depolarization effects.  The major benefit of such an experimental setup is that the irreducible SM background from the three-body Michel decay is reduced at the maximal positron momentum $p_{e^+}=m_\mu/2$ in the forward region (in the direction opposite  to the polarization of $\mu^+$). Since the SM decay amplitude is controlled by left-handed couplings, it vanishes for an exactly forward positron, if produced from a muon that is completely polarized.  For a highly polarized muon beam the SM background from $\mu^+\to e^+\nu \bar{\nu}$ is strongly suppressed in this part of the phase space, while the $\mu^+\to e^+a$ decay is allowed  for an LFV ALP with nonzero right-handed couplings to the SM leptons. 
MEGII-fwd can thus be used to search for an effectively massles ALP produced in $\mu^+\to e^+ a$, unless its couplings are aligned to the SM, i.e. mainly left-handed (${\CC{\rX}{ij} = - \CC{\rX5}{ij}} $). The final reach of {MEGII-fwd} depends on how well depolarization effects can be controlled, on the positron momentum resolution of the forward calorimeter, and on whether or not magnetic focusing is applied in order to increase the positron luminosity in the forward direction. In Table~\ref{expfut} we show the expected (optimistic) limit for the isotropic decay ${\rm BR}_{\rm proj} (\mu \to e X)_{\rm iso} = 7 \times 10^{-8}$~\cite{Calibbi:2020jvd}. Another strategy for MEG-II that does not require additional hardware is the MEGII-ALP proposal for $\mu \to e a \gamma$~\cite{Jho:2022snj}, which employs an alternative data-taking strategy that greatly increases the signal acceptance by adjusting the trigger selection while reducing the beam intensity. Indeed the standard MEGII trigger is optimized for the $\mu^+ \to e^+ \gamma$ decay, requiring the positron and photon to be back-to-back with equal energies $E_{e} = E_\gamma = m_\mu/2$. As a consequence, the trigger is suboptimal to probe $\mu \to e a \gamma$, where the signal rate peaks for a soft photon collinear with the positron. Implementing a new  trigger that selects events in this kinematic region would allow MEGII to search for $\mu^+\to e^+ a \gamma$ decays, giving a prospective limit on the decay rate that exceeds the one of the Crystal Ball collaboration~\cite{Bolton:1988af} by more than two orders of magnitude~\cite{Jho:2022snj}. 

Finally, it has been proposed to use detector validation datasets of Mu2e to conduct searches for $\mu^+ \to e^+ X$ decays at rest, as suggested in Ref.~\cite{Huang:2022xii, Hill:2023dym} under the shorthand Mu2e-X. An analogous search (COMET-X) was proposed for the COMET experiment  with $\mu^- \to e^- X$ decaying in orbit~\cite{Xing:2022rob}. The projected limit at both proposals is  ${\rm BR}_{\rm proj} (\mu \to e X)_{\rm iso} = 1 \times 10^{-7}$, and thus similar to the Mu3e reach. However, for massless dark bosons, both Mu2e and COMET proposals are likely to encounter similar systematic uncertainties as the $\mu \to e X$ search at Mu3e~\cite{Knapen:2023zgi}.


\section{ASTROPHYSICAL  CONSTRAINTS}
\label{AC}

It is remarkable that flavored dark sectors can also be probed with observations of core-collapse supernovae, which feature temperatures and densities high enough to sustain a sizable population of moderately heavy flavors. Muons and hyperons are, indeed, expected to emerge in the hot and dense proto-neutron star (PNS) formed during SN explosions~\cite{1960SvA,Oertel:2016bki,Bollig:2017lki,Fischer:2020vie}, and could decay into light dark  particles carrying energy away from the PNS. This new cooling mechanism can be constrained by observations of SN 1987A~\cite{MartinCamalich:2020dfe,Camalich:2020wac}. Specifically, the duration of the neutrino pulse would have been shorter than observed if the dark luminosity had been comparable to that of the neutrinos~\cite{Raffelt:1996wa}. These bounds have been extensively explored for muons in~\cite{Bollig:2020xdr,Calibbi:2020jvd,Croon:2020lrf, Caputo:2021rux,Manzari:2023gkt} and for hyperons in \cite{MartinCamalich:2020dfe,Camalich:2020wac,Alonso-Alvarez:2021oaj,Cavan-Piton:2024ayu,Fischer:2020vie}. 

In the case of hyperons, various models have been studied that induce flavor-changing neutral currents such as $\Lambda \to n X$. Focusing on the emission of light-dark bosons, the energy-loss rate per unit volume $Q$ can be estimated as the product of the hyperon number density $n_\Lambda$, the energy released per decay, approximately given by the hyperon-neutron mass difference, and the hyperon decay rate, giving $Q\propto n_\Lambda(m_\Lambda-m_n)\Gamma(\Lambda\to nX)$. 
One can further approximately describe the effects of neutron degeneracy\cite{Raffelt:1996wa,MartinCamalich:2020dfe} by a single number $F_n$, given by the thermal average of the Pauli blocking factor~\cite{Raffelt:1996wa}. An estimate of the  emissivity (the energy-loss rate per unit mass) $\epsilon$ in the nonrelativistic limit is, then,
\begin{equation}
\label{eq:SNhyp}
\epsilon \approx F_n Y_\Lambda \frac{m_\Lambda - m_n}{m_n}  \frac{ {\rm BR}(\Lambda \to n X)}{\tau_\Lambda} \, , 
\label{eq:emissivity_proxy}
\end{equation}
where $\tau_\Lambda$ is the hyperon lifetime and  $Y_\Lambda = n_\Lambda/n_B$ is the abundance of hyperons in the PNS, normalized by the baryonic number density. Adopting the classical upper limit on emissivity, $\epsilon_{\rm max} = 10^{19} \, {\rm erg \, s^{-1} \, g^{-1}}$~\cite{Raffelt:1996wa}, from SN~1987A under the conditions predicted for the PNS around 1 second post-bounce, one obtains
\begin{equation}
\label{eq:SNhypapp}
\epsilon \approx \epsilon_{\rm max} \left(\frac{Y_\Lambda}{0.01}\right)\left(\frac{0.7}{F_n}\right)\left(\frac{2 \times 10^{-9}}{{\rm BR}(\Lambda \to n X)}\right)~.
\end{equation}
Therefore, for a typical $\Lambda$ abundance of 1\%, this gives an upper limit on the branching fraction in the range of $10^{-9}$, which is many orders of magnitude more stringent than those obtained from laboratory searches, for example with the full data set at BESIII   (see Section~\ref{qfut}). 

Furthermore, as discussed in~\cite{Camalich:2020wac}, excessive emission of dark boson from hyperon decays would still occur in the deep-trapping regime (corresponding to large couplings or large branching fractions), as the dark luminosity would have to originate from the  surface where $\Lambda$'s remain in equilibrium with the plasma, which corresponds to a high-temperature region.

The estimate in Eq.~\ref{eq:SNhypapp} has been improved with a complete calculation of the decay process using kinetic theory, including the effects of the medium on the dispersion relation of the baryons~\cite{Camalich:2020wac}. Moreover, radial profiles of all the thermodynamical quantities relevant for the calculations of the rates can be extracted from spherically symmetric simulations, such as those reported in~\cite{Bollig:2020xdr}, designed for these types of studies. These simulations did not include hyperons as an ingredient in the nuclear equation of state or the feedback of the energy loss caused by the invisible decays of the $\Lambda$ in the SN simulation. Nonetheless, upper limits on the branching ratio were derived in Ref.~\cite{Camalich:2020wac}, with thermodynamic quantities at 1 second post-bounce recalculated using interpolation tables based on hyperonic extensions of the EOS used in the simulations. The upper limit that is obtained is,
\begin{equation}
\label{eq:SNhypFin}
{\rm BR}(\Lambda\to n X)\lesssim 8\times10^{-9},
\end{equation}
which is the weakest one among those obtained using the different simulations reported in Ref.~\cite{Bollig:2020xdr}. This result is a factor $\approx4$ weaker than the approximate expression in Eq.~\ref{eq:SNhypapp}, translating into a factor $\approx2$ weaker bound on the UV scale of the dark boson. For the case of scalar couplings, these bounds are $\Lambda_{\rm eff}\gtrsim7\times10^9$ GeV and $\Lambda_{\rm eff}\gtrsim5\times10^9$ GeV for $\CC{\rm S}{sd}$ and $\CC{\rm S5}{sd}$, respectively, and $\Lambda_{\rm eff}\gtrsim1\times 10^{10}$ GeV for the massless vector with dipole couplings.

Ref.~\cite{Fischer:2024ivh} reported the first simulations incorporating a hyperonic equation of state and energy losses due to invisible $\Lambda$ decays. These simulations demonstrate that such decays accelerate deleptonization of the PNS and enhance cooling, reducing the neutrino emission timescale by a factor of two and thereby validating previous analyses.
  
A similar analysis can be performed for the LFV muon decays $\mu\to e X$. Using the same approximations as for hyperons, one would find for this process an equation analogous to Eq.~\ref{eq:SNhyp} with the appropriate replacements, i.e. $\delta \to(m_\mu-m_e)/m_n$. Thus, for muon decays one obtains,
\begin{equation}
\label{eq:SNmuapp}
\epsilon \approx \epsilon_{\rm max} \left(\frac{Y_\mu}{0.03}\right)\left(\frac{0.5}{F_n}\right)\left(\frac{10^{-5}}{{\rm BR}(\mu \to e X)}\right), 
\end{equation}
which, contrary to the case of the quark couplings and the hyperon decays, is a bound much weaker than the ones obtained from laboratory experiments (see Ref.~\cite{Zhang:2023vva} for a related analysis using different production processes but obtaining similar limits).

We finally comment on astrophysical limits on flavor-diagonal couplings to fermions, which can be compared to the limits on flavor-violation. Writing  the scalar couplings to electrons and neutrons as 
${\cal L} = -a/\Lambda  (m_e \overline{e} (\mathbb{C}^{\rm S}_{ee} + \mathbb{C}^{\rm S5}_{ee} \gamma_5 ) e  + m_N \overline{N} (\mathbb{C}^{\rm S}_{NN} + \mathbb{C}^{\rm S5}_{NN} \gamma_5 ) N $, star cooling limits from Red Giants (RG) give $\Lambda/\mathbb{C}^{\rm S}_{ee,NN}\gtrsim 7 \times 10^{11} \GeV$~\cite{Hardy:2016kme}, while a recent analysis of the White Dwarf (WD) luminosity function gives $\Lambda/\mathbb{C}^{\rm S}_{ee,NN}\gtrsim 1 \times 10^{12} \GeV$~\cite{Bottaro:2023gep}, with only small differences between electron and nucleons. Limits on couplings to pseudoscalar currents are weaker, with $\Lambda/\mathbb{C}^{\rm S5}_{ee}\gtrsim 2 \times 10^{9} \GeV$ from WDs~\cite{MillerBertolami:2014rka} and $\Lambda/\mathbb{C}^{\rm S5}_{NN}\gtrsim 8 \times 10^{8} \GeV$  from SN~1987A~\cite{Carenza:2019pxu}. Note that for light vectors with diagonal couplings of the form in Eq.~\ref{vector} and derivatively coupled scalars (Eq.~\ref{axion}) the latter bounds apply. Even the most stringent limits on scalar couplings are surpassed by laboratory limits on $s \to d$ transitions, with potential to be extended in the near future by NA62. On the other hand star cooling limits on pseudoscalar diagonal couplings to nucleons and electrons (including couplings of light vectors and derivatively coupled axions) are at the level of $10^9 \GeV$, which is comparable to present limits on $\mu \to e$ transitions and prospects on $b \to s$ and $b \to d$ transitions. We emphasize that the bounds from star cooling of RGs (WDs) are not valid for dark boson masses above the typical core temperatures of about 10 keV (1 keV), while the laboratory limits discussed in the previous sections extend to much larger masses.


\section{COSMOLOGICAL  CONSTRAINTS}
\label{cosmo}

Since DM is our main motivation to look for missing energy in flavor-violating decays, we should also discuss their impact in cosmological scenarios where the dark boson is directly connected to the DM abundance. In the following we discuss these aspects in more detail. The easiest possibility is that the boson is stable even on cosmological scales, and is produced in the early universe in quantities that amount to the observed relic DM abundance of $\Omega_{\rm DM} h^2 = 0.12$. The preferred production mechanism depends crucially on the dark boson mass: very light bosons with masses much below the keV scale have to be non-thermally produced in order to avoid constraints on warm DM (WDM).  A classic production mechanism of this kind is misalignment, which has originally been proposed for the QCD axion~\cite{Preskill:1982cy, Abbott:1982af, Dine:1982ah}, and then generalized for general light feebly interacting bosons in Ref.~\cite{Arias:2012az}. However, a thermal population of such light particles can also be produced by the flavor-violating interactions in Eq.~\ref{setup}, which are then subject to stringent  constraints on dark radiation and WDM. If this production channel dominates over misalignment, it can fully account for the observed DM abundance via thermal freeze-in~\cite{Hall:2009bx}. This allows for a direct link between the size of flavor-violating decay rates and the DM relic abundance.

\subsection{Constraints from Dark Radiation}

While misalignment can easily yield light bosons behaving as cold DM in the required quantitites, such particles can be produced also thermally by direct couplings to SM particles and are relativistic at least at the early stage of the evolution of the Universe~\cite{Chang:1993gm}. In the context of  flavor-violating interactions in Eq.~\ref{setup} the main production channel would be through decays of SM fermions. For sufficiently large  couplings (sufficiently small $\Lambda/\mathbb{C}_{ij}$), such decays (and their inverse processes) bring the light bosons into thermal equilibrium with the SM plasma. When the temperature drops, the rate of inverse processes peters out, and the dark bosons maintain their freeze-out abundance when they decouple from the thermal bath. Alternatively, when the couplings are so small that thermal equilibrium is never achieved, a dark boson abundance slowly builds up from SM fermion decays, until the temperature drops below the mass of the mother fermion, so it becomes non-relativistic and its abundance is exponentially suppressed. 

Thermal relics are constrained by Big Bang Nucleosynthesis (BBN), observations of the Cosmic Microwave Background (CMB) and structure-formation. Very light (sub-eV) particles such as the QCD axion are  relativistic at recombination, and thus contribute to dark radiation (e.g. to the energy density stored in relativistic degrees of freedom), which is conveniently parameterized in terms of the effective number of additional neutrino species $\Delta N_{\rm eff}$. This observable is constrained by CMB observations and baryon acoustic oscillations (BAO), and the most recent combined analysis by Planck collaboration sets the upper bound $\Delta N_{\rm eff} \le 0.3$ at 95\% CL~\cite{Planck:2018vyg}. This bound is expected to be improved to 0.1 at the Simons Observatory~\cite{SimonsObservatory:2018koc} and eventually to 0.05 by CMB-S4 experiments~\cite{CMB-S4:2016ple}. 

These limits allow to constrain flavor-violating couplings of very light bosons, provided they are stable not only on collider but also on cosmological scales, which requires a calculation of their energy density at the time of recombination. The resulting constraints have been studied in Refs.~\cite{Baumann:2016wac, DEramo:2018vss, Arias-Aragon:2020shv, DEramo:2021usm, Badziak:2024szg}, and we report their results in the following. 

Starting with LFV transitions, present laboratory limits on  muon decays are stronger than the projected CMB-S4  bound from cosmology, giving $\Lambda_{\rm eff} \gtrsim 10^9 \GeV$ for $\mu \to e$ transitions~\cite{DEramo:2021usm}. Instead for $\tau \to \ell$ transitions (there is essentially no difference between $\ell = \mu,e$) the projected limit is of the order of  $10^8 \GeV$~\cite{Badziak:2024szg}, while the present bound from Planck cannot compete with the present Belle~II limit (according to Ref.~\cite{Badziak:2024szg}, which refined the calculation in Ref.~\cite{DEramo:2021usm} that suggested the opposite conclusion). 

In the quark sector, the NA62 limits on $s \to d$ exceed by far even future cosmology projections~\cite{DEramo:2021usm}, while CMB-S4 projections for $b \to s,d$ transitions are of the order of $\Lambda_{\rm eff} \gtrsim 10^8 \GeV$~\cite{Arias-Aragon:2020shv}, which interestingly are in the same ballpark as the projected limits for Belle~II. Note that Ref.~\cite{DEramo:2021usm} found more stringent cosmology limits, however it is not entirely clear whether the production rates from flavor-violating scattering processes were appropriately treated, due to complications from IR divergencies that can potentially give large unphysical enhancement factors (cf. Ref.~\cite{Aghaie:2024jkj}). For the same reason their projection on $c \to u$ transitions might be too optimistic, and a more conservative estimate can be obtained using the result in  Ref.~\cite{Baumann:2016wac} (which takes  into account only decays) finding $\Lambda_{\rm eff} \gtrsim 10^8 \GeV$, which again is at the same level of BESIII and STCF projections. We finally emphasize that the these limits from $\Delta N_{\rm eff}$ are invalidated if i) the dark boson is sufficiently heavy to avoid constraints from dark radiation and WDM, i.e. for roughly $m_X \gtrsim {\rm few}  \keV$; ii) if the boson is only stable on collider, but not on cosmological scales; or iii) if sizable couplings to SM particles substantially alter its thermal history.

\begin{figure}[h]
\includegraphics[width=3in]{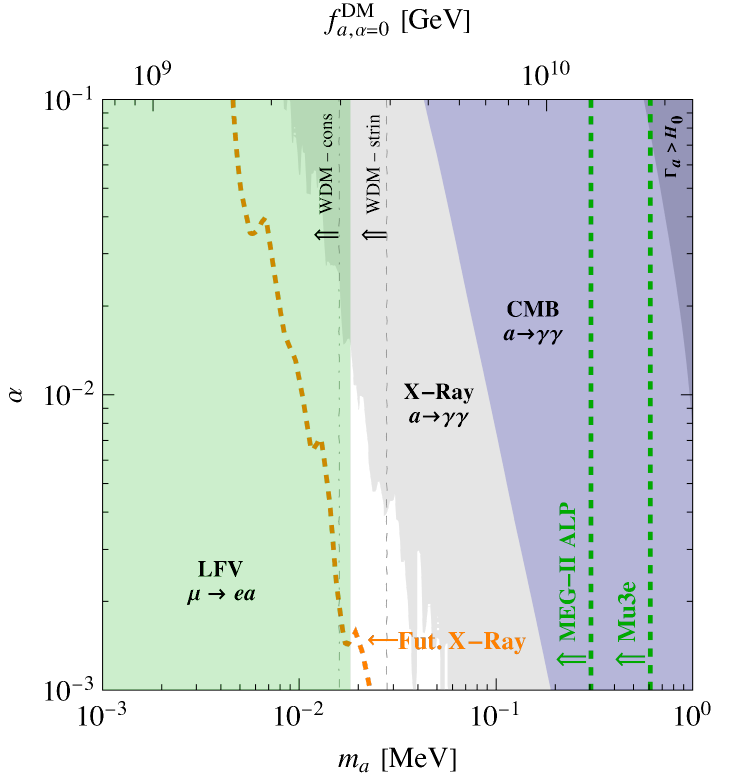}
\caption{ Allowed parameter space for DM freeze-in through LFV decays, taken from Ref.~\cite{Panci:2022wlc}. The ALP decay constant $f_a$ (top x-axis) is determined by requiring that the DM abundance today is produced through freeze-in, once  ALP mass $m_a$ (bottom x-axis) and the mixing angle  $\alpha$ is fixed, which is defined in Eq.~\ref{alphadef} (we choose the reference values $\alpha = 0$). The {\bf dark blue shaded}, {\bf blue shaded} and {\bf gray shaded} regions are excluded by the DM lifetime, CMB and X-ray constraints on decaying DM, respectively.  The reach of future X-rays searches is shown by {\bf dashed orange} lines.  Conservative (stringent) constraints on WDM requiring $m_{\rm WDM} \gtrsim 3.5 (5.3) \, \keV$ are shown as {\bf dotted-dashed}  ({\bf dashed})  gray lines. Present bound from  $\mu \to e a$ searches  are shown as {\bf green shaded} regions, while the prospects for future proposed searches at MEG II~\cite{Jho:2022snj} and Mu3e~\cite{Calibbi:2020jvd,Perrevoort:2018ttp} are shown as  {\bf dashed green} lines.
}
\label{money3}
\end{figure}

\subsection{DM Relic Abundance}
\label{LFVDM}

For dark bosons  with masses above few keV the $\Delta N_{\rm eff}$ and WDM bounds can be avoided, and one can consider scenarios where the flavor-violating decays are in fact the dominant production of light bosonic DM, reproducing the observed DM abundance via thermal freeze-in~\cite{Hall:2009bx}. The relic abundance is then set by the product of DM mass $m_X$ and flavor-violating decay rate, $\Omega_X h^2 \propto m_X \Gamma (f_i \to f_j X) \propto m_X (\CC{X,X5}{ij}/\Lambda)^2$, giving a prediction for the decay rate as a function of the dark boson mass, which otherwise requires extensive flavor model-building (cf. Sec.~\ref{setmot}). This kind of scenarios serve as one possible motivation for explicit experimental targets, directly connecting the flavor-violating decays to the DM abundance. Therefore, we give some details of these models below, focusing  on the case of axion-like particles discussed in Refs.~\cite{Panci:2022wlc, Aghaie:2024jkj}. 

The main challenge of this scenario is DM stability, as the ALP can always decay into two photons, with a decay rate set by 
\begin{equation}
\Gamma_{a \to \gamma \gamma} \approx \frac{\alpha_{\rm em}^2}{64 \pi^3} \frac{m_a^3}{f_a^2}  \left| C_\gamma \right|^2 \, , \qquad \qquad C_\gamma = E- 1.92 N +\sum_i C^A_{ii} \frac{m_a^2}{12 m_{i}^2}  \, , 
  \end{equation}
up to higher powers of $m_a^2/m_i^2$, and where the sum runs over all electrically charged SM fermions. Here $C^A_{ii}$ are the diagonal axion couplings to fermions in Eq.~\ref{axion} and $E$($N$) are the electromagnetic (color) anomaly coefficients of the model, determined by the fermion couplings. Although it is relatively easy to make the axion lifetime larger than the age of the universe, $1/H_0 \sim 10^{17} {\rm sec}$, it turns out that in the relevant axion mass range much stronger constraints on the decay rate arise from X-ray telescopes, requiring $\Gamma_{a \to \gamma \gamma} \gtrsim 10^{28} {\rm sec}$~\cite{Panci:2022wlc}. This implies that the underlying PQ symmetry must be anomaly-free, and one needs $m_a \ll m_i$ and/or $C_{ii}^A \ll C^{A,V}_{ij}$. In fact, note that DM production is controlled by $C_{i \ne j}^{A,V}$ while DM stability is governed by $C^A_{ii}$, so in principle these are independent parameters, although a strong hierarchy  $C^A_{ii} \ll C^{A,V}_{ij}$ seems unnatural. In any case, ALP stability gives an upper bound on $m_a$, while a lower bound arises from experimental limits on the decay rate (since $\Omega_a h^2 \propto m_a \Gamma (f_i \to f_j a$)) and from Lyman-$\alpha$ limits on WDM, which in the case of freeze-in production require $m_a \gtrsim 10 \keV$.

As an explicit example we consider the effective 2-flavor scenario discussed in Ref.~\cite{Panci:2022wlc}, where two right-handed SM leptons are oppositely charged under PQ and a single angle parametrizes the rotation to the mass basis. This simplified model has just three parameters: the axion mass $m_a$, the axion decay constant $f_a$, and a single angle $0 \le \alpha \le \pi/2$, which controls the couplings to leptons, e.g. for the $\mu e$ scenario 
\begin{equation}
C^A_{ee} = - C^A_{\mu \mu} = \sin \alpha \, , \qquad \qquad C_{ \mu e}^A = C_{ \mu e}^V = \cos \alpha \, .    
\label{alphadef}
\end{equation}
After fixing the axion decay constant to reproduce the observed abundance, one is left with a two-dimensional parameter space, where constraints on decaying DM, WDM and direct laboratory searches are imposed, see Fig.~\ref{money3}. The stringent bounds from X-ray searches require diagonal axion couplings to be smaller than off-diagonal ones by at least 1\%, ruling out the limit of exact flavor conservation $\alpha = \pi/2$. Present limits from laboratory searches are roughly at the same level as WDM constraints, but the proposed searches at
MEG~II and Mu3e will probe almost the entire allowed parameter space. These experiments have the unique opportunity of probing directly the very same decay that could have produced axion DM in the early Universe.


\section{SUMMARY AND OUTLOOK}

\begin{summary}[SUMMARY POINTS]
\begin{enumerate}
\item Flavor-violating decays with light invisible states, mimicking SM three-body decays with neutrino pairs, can be probed at colliders. In contrast to heavy new physics contributions, these searches are controlled by dimension-five operators, providing exceptional sensitivity to high-energy scales, with current limits ranging from $10^7$~GeV (tau and charm) to nearly $10^{12}$~GeV (kaons).
\item Light dark bosons with flavor-violating couplings have decent theoretical motivation by the strong CP problem, the dark matter relic abundance and/or the SM flavor puzzle, for example as QCD axions of a flavor non-universal PQ symmetry.  
\item Muons and hyperons are present in proto-neutron stars formed during core-collapse supernovae. Their decays into invisible states introduce a new cooling mechanism  constrained by SN~1987A. The resulting limits on hyperon decays are stronger than those from laboratory experiments by several orders of magnitude.
\item  Flavor-violating decays can take place in the early universe producing a relic abundance of light dark states. While DM masses below a few keV are constrained by dark radiation and warm DM limits, heavier bosons can potentially account for the observed DM abundance with a direct link to the size of flavor-violating couplings. 
\end{enumerate}
\end{summary}

\begin{issues}[FUTURE ISSUES]
\begin{enumerate}
\item Flavor constraints on light dark  sectors can be strengthened considerably by improving existing searches with larger data sets, applying dedicated analysis strategies to current data, or conducting entirely new searches. 
\item Projected sensitivities to the UV scale  reach up to more than $10^{12}~\GeV$ in kaon decays (NA62), $10^{10}~\GeV$ in muon decays (Mu3e/MEG) and $10^8~\GeV$ in $B$-meson (Belle II) and charmed-hadron decays (BESIII/STCF).  
\item Predictions of flavor-violating couplings in explicit models are rare. Their determination in motivated SM extensions will prove valuable to define explicit experimental targets for probing dark flavored sectors.
\item In many cases there is an interesting interplay between future collider and cosmological constraints, which motivates a more precise assessment of prospective collider limits and  DM production rates in the early universe.
\end{enumerate}
\end{issues}


\section*{DISCLOSURE STATEMENT}
The authors are not aware of any affiliations, memberships, funding, or financial holdings that
might be perceived as affecting the objectivity of this review. 

\section*{ACKNOWLEDGMENTS}
JMC thanks MICINN for funding through the grant ``DarkMaps'' PID2022-142142NB-I00 and from the European Union through the grant ``UNDARK'' of the Widening participation and spreading excellence programme (project number 101159929). This work has received support from the European
Union's Horizon 2020 research and innovation programme under the Marie Sk{\l}odowska-Curie grant
agreement No 860881-HIDDeN and is partially supported by project B3a and C3b of the DFG-funded
Collaborative Research Center TRR257 ``Particle Physics Phenomenology after the Higgs Discovery". RZ thanks the INFN National Laboratory of Frascati for kind hospitality, where this work has been partially completed.

\bibliographystyle{ar-style5}

\bibliography{references}

\begin{thebibliography}{165}
\expandafter\ifx\csname natexlab\endcsname\relax\def\natexlab#1{#1}\fi

\bibitem{Knapen:2017xzo}
Knapen S, Lin T, Zurek KM.
\newblock \textit{Phys. Rev. D} 96(11):115021 (2017)

\bibitem{Lin:2019uvt}
Lin T.
\newblock \textit{PoS} 333:009 (2019)

\bibitem{Zurek:2024qfm}
Zurek KM.
\newblock \textit{Ann. Rev. Nucl. Part. Sci.} 74:287--319 (2024)

\bibitem{Cirelli:2024ssz}
Cirelli M, Strumia A, Zupan J  (2024)

\bibitem{Wilczek:1982rv}
Wilczek F.
\newblock \textit{Phys. Rev. Lett.} 49:1549--1552 (1982)

\bibitem{Holdom:1985ag}
Holdom B.
\newblock \textit{Phys. Lett. B} 166:196--198 (1986)

\bibitem{Dobrescu:2004wz}
Dobrescu BA.
\newblock \textit{Phys. Rev. Lett.} 94:151802 (2005)

\bibitem{Elor:2018twp}
Elor G, Escudero M, Nelson A.
\newblock \textit{Phys. Rev. D} 99(3):035031 (2019)

\bibitem{Kamenik:2011vy}
Kamenik JF, Smith C.
\newblock \textit{JHEP} 03:090 (2012)

\bibitem{LHCb:2018roe}
Aaij R, et~al.  (2018)

\bibitem{Belle-II:2018jsg}
Altmannshofer W, et~al.
\newblock \textit{PTEP} 2019(12):123C01 (2019), [Erratum: PTEP 2020, 029201
  (2020)]

\bibitem{BESIII:2020nme}
Ablikim M, et~al.
\newblock \textit{Chin. Phys. C} 44(4):040001 (2020)

\bibitem{NA62:2017rwk}
Cortina~Gil E, et~al.
\newblock \textit{JINST} 12(05):P05025 (2017)

\bibitem{Yamanaka:2012yma}
Yamanaka T.
\newblock \textit{PTEP} 2012:02B006 (2012)

\bibitem{Baldini:2018nnn}
Baldini AM, et~al.
\newblock \textit{Eur. Phys. J. C} 78(5):380 (2018)

\bibitem{Eguren:2024oov}
Eguren JF, Klingel S, Stamou E, Tabet M, Ziegler R.
\newblock \textit{JHEP} 08:111 (2024)

\bibitem{Fabbrichesi:2017vma}
Fabbrichesi M, Gabrielli E, Mele B.
\newblock \textit{Phys. Rev. Lett.} 119(3):031801 (2017)

\bibitem{DiLuzio:2020wdo}
Di~Luzio L, Giannotti M, Nardi E, Visinelli L.
\newblock \textit{Phys. Rept.} 870:1--117 (2020)

\bibitem{Feruglio:2015jfa}
Feruglio F.
\newblock \textit{Eur. Phys. J. C} 75(8):373 (2015)

\bibitem{Wilczek:1977pj}
Wilczek F.
\newblock \textit{Phys. Rev. Lett.} 40:279--282 (1978)

\bibitem{Weinberg:1977ma}
Weinberg S.
\newblock \textit{Phys. Rev. Lett.} 40:223--226 (1978)

\bibitem{Peccei:1977hh}
Peccei RD, Quinn HR.
\newblock \textit{Phys. Rev. Lett.} 38:1440--1443 (1977)

\bibitem{Peccei:1977ur}
Peccei RD, Quinn HR.
\newblock \textit{Phys. Rev. D} 16:1791--1797 (1977)

\bibitem{GrillidiCortona:2015jxo}
Grilli~di Cortona G, Hardy E, Pardo~Vega J, Villadoro G.
\newblock \textit{JHEP} 01:034 (2016)

\bibitem{Preskill:1982cy}
Preskill J, Wise MB, Wilczek F.
\newblock \textit{Phys. Lett. B} 120:127--132 (1983)

\bibitem{Abbott:1982af}
Abbott LF, Sikivie P.
\newblock \textit{Phys. Lett. B} 120:133--136 (1983)

\bibitem{Dine:1982ah}
Dine M, Fischler W.
\newblock \textit{Phys. Lett. B} 120:137--141 (1983)

\bibitem{MartinCamalich:2020dfe}
Martin~Camalich J, Pospelov M, Vuong PNH, Ziegler R, Zupan J.
\newblock \textit{Phys. Rev. D} 102(1):015023 (2020)

\bibitem{Kim:1979if}
Kim JE.
\newblock \textit{Phys. Rev. Lett.} 43:103 (1979)

\bibitem{Shifman:1979if}
Shifman MA, Vainshtein AI, Zakharov VI.
\newblock \textit{Nucl. Phys. B} 166:493--506 (1980)

\bibitem{Dine:1981rt}
Dine M, Fischler W, Srednicki M.
\newblock \textit{Phys. Lett. B} 104:199--202 (1981)

\bibitem{Zhitnitsky:1980tq}
Zhitnitsky AR.
\newblock \textit{Sov. J. Nucl. Phys.} 31:260 (1980)

\bibitem{Bardeen:1986yb}
Bardeen WA, Peccei RD, Yanagida T.
\newblock \textit{Nucl. Phys. B} 279:401--428 (1987)

\bibitem{Geng:1988nc}
Geng CQ, Ng JN.
\newblock \textit{Phys. Rev. D} 39:1449 (1989)

\bibitem{Hindmarsh:1997ac}
Hindmarsh M, Moulatsiotis P.
\newblock \textit{Phys. Rev. D} 56:8074--8081 (1997)

\bibitem{DiLuzio:2017ogq}
Di~Luzio L, Mescia F, Nardi E, Panci P, Ziegler R.
\newblock \textit{Phys. Rev. Lett.} 120(26):261803 (2018)

\bibitem{Bjorkeroth:2019jtx}
Bj\"orkeroth F, Di~Luzio L, Mescia F, Nardi E, Panci P, Ziegler R.
\newblock \textit{Phys. Rev. D} 101(3):035027 (2020)

\bibitem{Saikawa:2019lng}
Saikawa K, Yanagida TT.
\newblock \textit{JCAP} 03:007 (2020)

\bibitem{Badziak:2021apn}
Badziak M, Grilli~di Cortona G, Tabet M, Ziegler R.
\newblock \textit{JHEP} 10:181 (2021)

\bibitem{Davidson:1981zd}
Davidson A, Wali KC.
\newblock \textit{Phys. Rev. Lett.} 48:11 (1982)

\bibitem{Berezhiani:1989fp}
Berezhiani ZG, Khlopov MY.
\newblock \textit{Z. Phys. C} 49:73--78 (1991)

\bibitem{Ema:2016ops}
Ema Y, Hamaguchi K, Moroi T, Nakayama K.
\newblock \textit{JHEP} 01:096 (2017)

\bibitem{Calibbi:2016hwq}
Calibbi L, Goertz F, Redigolo D, Ziegler R, Zupan J.
\newblock \textit{Phys. Rev. D} 95(9):095009 (2017)

\bibitem{Froggatt:1978nt}
Froggatt CD, Nielsen HB.
\newblock \textit{Nucl. Phys. B} 147:277--298 (1979)

\bibitem{Ibanez:1994ig}
Ibanez LE, Ross GG.
\newblock \textit{Phys. Lett. B} 332:100--110 (1994)

\bibitem{Binetruy:1994ru}
Binetruy P, Ramond P.
\newblock \textit{Phys. Lett. B} 350:49--57 (1995)

\bibitem{Linster:2018avp}
Linster M, Ziegler R.
\newblock \textit{JHEP} 08:058 (2018)

\bibitem{Calibbi:2020jvd}
Calibbi L, Redigolo D, Ziegler R, Zupan J.
\newblock \textit{JHEP} 09:173 (2021)

\bibitem{Choi:2017gpf}
Choi K, Im SH, Park CB, Yun S.
\newblock \textit{JHEP} 11:070 (2017)

\bibitem{Chala:2020wvs}
Chala M, Guedes G, Ramos M, Santiago J.
\newblock \textit{Eur. Phys. J. C} 81(2):181 (2021)

\bibitem{Bauer:2021mvw}
Bauer M, Neubert M, Renner S, Schnubel M, Thamm A.
\newblock \textit{JHEP} 09:056 (2022)

\bibitem{Hall:2009bx}
Hall LJ, Jedamzik K, March-Russell J, West SM.
\newblock \textit{JHEP} 03:080 (2010)

\bibitem{Co:2017mop}
Co RT, Hall LJ, Harigaya K.
\newblock \textit{Phys. Rev. Lett.} 120(21):211602 (2018)

\bibitem{Greljo:2024evt}
Greljo A, Smolkovi\v{c} A, Valenti A.
\newblock \textit{JHEP} 09:174 (2024)

\bibitem{Heeck:2020nbq}
Heeck J.
\newblock \textit{Phys. Lett. B} 813:136043 (2021)

\bibitem{Chikashige:1980ui}
Chikashige Y, Mohapatra RN, Peccei RD.
\newblock \textit{Phys. Lett. B} 98:265--268 (1981)

\bibitem{Schechter:1981cv}
Schechter J, Valle JWF.
\newblock \textit{Phys. Rev. D} 25:774 (1982)

\bibitem{Cicoli:2013ana}
Cicoli M. 2013.
\newblock In \textit{{9th Patras Workshop on Axions, WIMPs and WISPs}}

\bibitem{Feruglio:2024dnc}
Feruglio F, Ziegler R  (2024)

\bibitem{Feruglio:2017spp}
Feruglio F (2019).
\newblock {Are neutrino masses modular forms?}
\newblock  227--266

\bibitem{Feruglio:2023uof}
Feruglio F, Strumia A, Titov A.
\newblock \textit{JHEP} 07:027 (2023)

\bibitem{Nelson:2011sf}
Nelson AE, Scholtz J.
\newblock \textit{Phys. Rev. D} 84:103501 (2011)

\bibitem{Arias:2012az}
Arias P, Cadamuro D, Goodsell M, Jaeckel J, Redondo J, Ringwald A.
\newblock \textit{JCAP} 06:013 (2012)

\bibitem{Smolkovic:2019jow}
Smolkovi\v{c} A, Tammaro M, Zupan J.
\newblock \textit{JHEP} 10:188 (2019), [Erratum: JHEP 02, 033 (2022)]

\bibitem{Bonnefoy:2019lsn}
Bonnefoy Q, Dudas E, Pokorski S.
\newblock \textit{JHEP} 01:191 (2020)

\bibitem{Greljo:2023bix}
Greljo A, Thomsen AE.
\newblock \textit{Eur. Phys. J. C} 84(2):213 (2024)

\bibitem{Foot:1994vd}
Foot R, He XG, Lew H, Volkas RR.
\newblock \textit{Phys. Rev. D} 50:4571--4580 (1994)

\bibitem{Ardu:2022zom}
Ardu M, Kirk F.
\newblock \textit{Eur. Phys. J. C} 83(5):394 (2023)

\bibitem{Grinstein:2010ve}
Grinstein B, Redi M, Villadoro G.
\newblock \textit{JHEP} 11:067 (2010)

\bibitem{Fabbrichesi:2020wbt}
Fabbrichesi M, Gabrielli E, Lanfranchi G  (2020)

\bibitem{Patt:2006fw}
Patt B, Wilczek F  (2006)

\bibitem{Batell:2009di}
Batell B, Pospelov M, Ritz A.
\newblock \textit{Phys. Rev. D} 80:095024 (2009)

\bibitem{Lanfranchi:2020crw}
Lanfranchi G, Pospelov M, Schuster P.
\newblock \textit{Ann. Rev. Nucl. Part. Sci.} 71:279--313 (2021)

\bibitem{Alekhin:2015byh}
Alekhin S, et~al.
\newblock \textit{Rept. Prog. Phys.} 79(12):124201 (2016)

\bibitem{SHiP:2015vad}
Anelli M, et~al.  (2015)

\bibitem{Feng:2017uoz}
Feng JL, Galon I, Kling F, Trojanowski S.
\newblock \textit{Phys. Rev. D} 97(3):035001 (2018)

\bibitem{FASER:2018bac}
Ariga A, et~al.  (2018)

\bibitem{Dobrich:2018jyi}
D\"obrich B, Ertas F, Kahlhoefer F, Spadaro T.
\newblock \textit{Phys. Lett. B} 790:537--544 (2019)

\bibitem{Balkin:2024qtf}
Balkin R, Burger N, Feng JL, Shadmi Y  (2024)

\bibitem{Feng:1997tn}
Feng JL, Moroi T, Murayama H, Schnapka E.
\newblock \textit{Phys. Rev. D} 57:5875--5892 (1998)

\bibitem{Bjorkeroth:2018dzu}
Bj\"orkeroth F, Chun EJ, King SF.
\newblock \textit{JHEP} 08:117 (2018)

\bibitem{Carmona:2021seb}
Carmona A, Scherb C, Schwaller P.
\newblock \textit{JHEP} 08:121 (2021)

\bibitem{Ibarra:2021xyk}
Ibarra A, Mar\'\i{}n M, Roig P.
\newblock \textit{Phys. Lett. B} 827:136933 (2022)

\bibitem{Heeck:2016xkh}
Heeck J.
\newblock \textit{Phys. Lett. B} 758:101--105 (2016)

\bibitem{Gabrielli:2016cut}
Gabrielli E, Mele B, Raidal M, Venturini E.
\newblock \textit{Phys. Rev. D} 94(11):115013 (2016)

\bibitem{Su:2019ipw}
Su JY, Tandean J.
\newblock \textit{Phys. Rev. D} 101(3):035044 (2020)

\bibitem{Su:2020xwt}
Su JY, Tandean J.
\newblock \textit{Eur. Phys. J. C} 80(9):824 (2020)

\bibitem{Camalich:2020wac}
Camalich JM, Terol-Calvo J, Tolos L, Ziegler R.
\newblock \textit{Phys. Rev. D} 103(12):L121301 (2021)

\bibitem{NA62:2021zjw}
Cortina~Gil E, et~al.
\newblock \textit{JHEP} 06:093 (2021)

\bibitem{BESIII:2023utd}
Ablikim M, et~al.
\newblock \textit{Phys. Lett. B} 852:138614 (2024)

\bibitem{BESIII:2022vrr}
Ablikim M, et~al.
\newblock \textit{Phys. Rev. D} 106(7):072008 (2022)

\bibitem{BaBar:2004xlo}
Aubert B, et~al.
\newblock \textit{Phys. Rev. Lett.} 94:101801 (2005)

\bibitem{ALEPH:2000vvi}
Barate R, et~al.
\newblock \textit{Eur. Phys. J. C} 19:213--227 (2001)

\bibitem{Alonso-Alvarez:2023mgc}
Alonso-\'Alvarez G, Escudero~Abenza M.
\newblock \textit{Eur. Phys. J. C} 84(5):553 (2024)

\bibitem{Belle-II:2023esi}
Adachi I, et~al.
\newblock \textit{Phys. Rev. D} 109(11):112006 (2024)

\bibitem{BaBar:2013npw}
Lees JP, et~al.
\newblock \textit{Phys. Rev. D} 87(11):112005 (2013)

\bibitem{Altmannshofer:2023hkn}
Altmannshofer W, Crivellin A, Haigh H, Inguglia G, Martin~Camalich J.
\newblock \textit{Phys. Rev. D} 109(7):075008 (2024)

\bibitem{Jodidio:1986mz}
Jodidio A, et~al.
\newblock \textit{Phys. Rev. D} 34:1967 (1986), [Erratum: Phys.Rev.D 37, 237
  (1988)]

\bibitem{Bayes:2014lxz}
Bayes R, et~al.
\newblock \textit{Phys. Rev. D} 91(5):052020 (2015)

\bibitem{Belle-II:2022heu}
Adachi I, et~al.
\newblock \textit{Phys. Rev. Lett.} 130(18):181803 (2023)

\bibitem{CLEO:2008ffk}
Eisenstein BI, et~al.
\newblock \textit{Phys. Rev. D} 78:052003 (2008)

\bibitem{Guadagnoli:2025xnt}
Guadagnoli D, Iohner A, Lazzeroni C, Martinez~Santos D, Swallow JC, Toni C
  (2025)

\bibitem{BESIII:2019vhn}
Ablikim M, et~al.
\newblock \textit{Phys. Rev. Lett.} 123(21):211802 (2019)

\bibitem{BESIII:2024vlt}
Ablikim M, et~al.  (2024)

\bibitem{Albrecht:2019zul}
Albrecht J, Stamou E, Ziegler R, Zwicky R.
\newblock \textit{JHEP} 21:139 (2020)

\bibitem{Adler:2000ic}
Adler S, et~al.
\newblock \textit{Phys. Rev. D} 63:032004 (2001)

\bibitem{Sadovsky:2023cxu}
Sadovsky AS, et~al.
\newblock \textit{Eur. Phys. J. C} 84(3):266 (2024)

\bibitem{E391a:2011aa}
Ogata R, et~al.
\newblock \textit{Phys. Rev. D} 84:052009 (2011)

\bibitem{Batley:2010zza}
Batley JR, et~al.
\newblock \textit{Eur. Phys. J. C} 70:635--657 (2010)

\bibitem{Batley:2012rf}
Batley JR, et~al.
\newblock \textit{Phys. Lett. B} 715:105--115 (2012), [Addendum: Phys.Lett.B
  740, 364--364 (2015)]

\bibitem{Littenberg:1995zy}
Littenberg LS, Valencia G.
\newblock \textit{Phys. Lett. B} 385:379--384 (1996)

\bibitem{Chiang:2000bg}
Chiang CW, Gilman FJ.
\newblock \textit{Phys. Rev. D} 62:094026 (2000)

\bibitem{Cavan-Piton:2024pqp}
Cavan-Piton M, Guadagnoli D, Iohner A, Martinez~Santos D, Vittorio L  (2024)

\bibitem{Bona:2024bue}
Bona M, et~al.
\newblock \textit{PoS} WIFAI2023:007 (2024)

\bibitem{Bolton:1988af}
Bolton RD, et~al.
\newblock \textit{Phys. Rev. D} 38:2077 (1988)

\bibitem{Perrevoort:2018ttp}
Perrevoort AK.
\newblock \textit{SciPost Phys. Proc.} 1:052 (2019)

\bibitem{Perrevoort:2018okj}
Perrevoort AK. 2018.
\newblock {Sensitivity Studies on New Physics in the Mu3e Experiment and
  Development of Firmware for the Front-End of the Mu3e Pixel Detector}.
\newblock Ph.D. thesis, U. Heidelberg (main)

\bibitem{Hill:2023dym}
Hill RJ, Plestid R, Zupan J.
\newblock \textit{Phys. Rev. D} 109(3):035025 (2024)

\bibitem{Badziak:2024szg}
Badziak M, Harigaya K, \L{}ukawski M, Ziegler R.
\newblock \textit{JHEP} 09:136 (2024)

\bibitem{Adler:2008zza}
Adler S, et~al.
\newblock \textit{Phys. Rev. D} 77:052003 (2008)

\bibitem{Goudzovski:2022vbt}
Goudzovski E, et~al.
\newblock \textit{Rept. Prog. Phys.} 86(1):016201 (2023)

\bibitem{Zhou:2021rgi}
Zhou X.
\newblock \textit{PoS} CHARM2020:007 (2021)

\bibitem{BESIII:2021slf}
Ablikim M, et~al.
\newblock \textit{Phys. Rev. D} 105(7):L071102 (2022)

\bibitem{Belle:2017oht}
Grygier J, et~al.
\newblock \textit{Phys. Rev. D} 96(9):091101 (2017), [Addendum: Phys.Rev.D 97,
  099902 (2018)]

\bibitem{Berger:2014vba}
Berger N.
\newblock \textit{Nucl. Phys. B Proc. Suppl.} 248-250:35--40 (2014)

\bibitem{Blondel:2013ia}
Blondel A, et~al.  (2013)

\bibitem{Adamov:2018vin}
Abramishvili R, et~al.
\newblock \textit{PTEP} 2020(3):033C01 (2020)

\bibitem{Bartoszek:2014mya}
Bartoszek L, et~al.  (2014)

\bibitem{Perez:2019cdy}
Rodr\'\i{}guez~P\'erez D. 2019.
\newblock In \textit{{17th Conference on Flavor Physics and CP Violation}}

\bibitem{Achasov:2023gey}
Achasov M, et~al.
\newblock \textit{Front. Phys. (Beijing)} 19(1):14701 (2024)

\bibitem{Guadagnoli:2021fcj}
Guadagnoli D, Park CB, Tenchini F.
\newblock \textit{Phys. Lett. B} 822:136701 (2021)

\bibitem{DeLaCruz-Burelo:2020ozf}
De~La Cruz-Burelo E, Hernandez-Villanueva M, De~Yta-Hernandez A.
\newblock \textit{Phys. Rev. D} 102(11):115001 (2020)

\bibitem{Knapen:2023zgi}
Knapen S, Langhoff K, Opferkuch T, Redigolo D  (2023)

\bibitem{Banerjee:2022nbr}
Banerjee P, Coutinho A, Engel T, Gurgone A, Signer A, Ulrich Y.
\newblock \textit{SciPost Phys.} 15(1):021 (2023)

\bibitem{Banerjee:2020rww}
Banerjee P, Engel T, Signer A, Ulrich Y.
\newblock \textit{SciPost Phys.} 9:027 (2020)

\bibitem{Jho:2022snj}
Jho Y, Knapen S, Redigolo D.
\newblock \textit{JHEP} 10:029 (2022)

\bibitem{Huang:2022xii}
Huang S. 2022.
\newblock {Search for Lepton Flavor Violation in Two Body Muon and Pion Decay
  at Rest}.
\newblock Ph.D. thesis, Purdue U., West Lafayette

\bibitem{Xing:2022rob}
Xing T, Wu C, Miao H, Li HB, Li W, et~al.
\newblock \textit{Chin. Phys. C} 47(1):013108 (2023)

\bibitem{1960SvA}
Ambartsumyan VA, Saakyan GS.
\newblock \textit{Soviet Astronomy} 4:187 (1960)

\bibitem{Oertel:2016bki}
Oertel M, Hempel M, Kl\"ahn T, Typel S.
\newblock \textit{Rev. Mod. Phys.} 89(1):015007 (2017)

\bibitem{Bollig:2017lki}
Bollig R, Janka HT, Lohs A, Martinez-Pinedo G, Horowitz CJ, Melson T.
\newblock \textit{Phys. Rev. Lett.} 119(24):242702 (2017)

\bibitem{Fischer:2020vie}
Fischer T, Guo G, Mart\'\i{}nez-Pinedo G, Liebend\"orfer M, Mezzacappa A.
\newblock \textit{Phys. Rev. D} 102(12):123001 (2020)

\bibitem{Raffelt:1996wa}
Raffelt GG (1996)

\bibitem{Bollig:2020xdr}
Bollig R, DeRocco W, Graham PW, Janka HT.
\newblock \textit{Phys. Rev. Lett.} 125(5):051104 (2020), [Erratum:
  Phys.Rev.Lett. 126, 189901 (2021)]

\bibitem{Croon:2020lrf}
Croon D, Elor G, Leane RK, McDermott SD.
\newblock \textit{JHEP} 01:107 (2021)

\bibitem{Caputo:2021rux}
Caputo A, Raffelt G, Vitagliano E.
\newblock \textit{Phys. Rev. D} 105(3):035022 (2022)

\bibitem{Manzari:2023gkt}
Manzari CA, Martin~Camalich J, Spinner J, Ziegler R.
\newblock \textit{Phys. Rev. D} 108(10):103020 (2023)

\bibitem{Alonso-Alvarez:2021oaj}
Alonso-\'Alvarez G, Elor G, Escudero M, Fornal B, Grinstein B, Martin~Camalich
  J.
\newblock \textit{Phys. Rev. D} 105(11):115005 (2022)

\bibitem{Cavan-Piton:2024ayu}
Cavan-Piton M, Guadagnoli D, Oertel M, Seong H, Vittorio L.
\newblock \textit{Phys. Rev. Lett.} 133(12):121002 (2024)

\bibitem{Fischer:2024ivh}
Fischer T, Martin~Camalich J, Kochankovski H, Tolos L  (2024)

\bibitem{Zhang:2023vva}
Zhang HY, Hagimoto R, Long AJ.
\newblock \textit{Phys. Rev. D} 109(10):103005 (2024)

\bibitem{Hardy:2016kme}
Hardy E, Lasenby R.
\newblock \textit{JHEP} 02:033 (2017)

\bibitem{Bottaro:2023gep}
Bottaro S, Caputo A, Raffelt G, Vitagliano E.
\newblock \textit{JCAP} 07:071 (2023)

\bibitem{MillerBertolami:2014rka}
Miller~Bertolami MM, Melendez BE, Althaus LG, Isern J.
\newblock \textit{JCAP} 10:069 (2014)

\bibitem{Carenza:2019pxu}
Carenza P, Fischer T, Giannotti M, Guo G, Mart\'\i{}nez-Pinedo G, Mirizzi A.
\newblock \textit{JCAP} 10(10):016 (2019), [Erratum: JCAP 05, E01 (2020)]

\bibitem{Chang:1993gm}
Chang S, Choi K.
\newblock \textit{Phys. Lett. B} 316:51--56 (1993)

\bibitem{Planck:2018vyg}
Aghanim N, et~al.
\newblock \textit{Astron. Astrophys.} 641:A6 (2020), [Erratum:
  Astron.Astrophys. 652, C4 (2021)]

\bibitem{SimonsObservatory:2018koc}
Ade P, et~al.
\newblock \textit{JCAP} 02:056 (2019)

\bibitem{CMB-S4:2016ple}
Abazajian KN, et~al.  (2016)

\bibitem{Baumann:2016wac}
Baumann D, Green D, Wallisch B.
\newblock \textit{Phys. Rev. Lett.} 117(17):171301 (2016)

\bibitem{DEramo:2018vss}
D'Eramo F, Ferreira RZ, Notari A, Bernal JL.
\newblock \textit{JCAP} 11:014 (2018)

\bibitem{Arias-Aragon:2020shv}
Arias-Arag\'on F, D'Eramo F, Ferreira RZ, Merlo L, Notari A.
\newblock \textit{JCAP} 03:090 (2021)

\bibitem{DEramo:2021usm}
D'Eramo F, Yun S.
\newblock \textit{Phys. Rev. D} 105(7):075002 (2022)

\bibitem{Aghaie:2024jkj}
Aghaie M, Armando G, Conaci A, Dondarini A, Matak P, et~al.
\newblock \textit{Phys. Lett. B} 856:138923 (2024)

\bibitem{Panci:2022wlc}
Panci P, Redigolo D, Schwetz T, Ziegler R.
\newblock \textit{Phys. Lett. B} 841:137919 (2023)

\end{thebibliography}

\end{document}